\documentclass[journal]{IEEEtran}
\usepackage{amssymb,amsfonts}
\usepackage{amsmath}
\usepackage{cite}
\usepackage{graphicx}
\usepackage{array}
\usepackage{multirow}
\usepackage{longtable}
\usepackage{rotating}
\usepackage{color}
\usepackage{algorithm}
\usepackage{algorithmic}
\usepackage{stfloats}
\usepackage{textcomp}
\usepackage[numbers,sort&compress]{natbib}
\usepackage{txfonts}
\usepackage{stfloats}

\ifCLASSINFOpdf
\else
\fi
\begin{document}
\title{Information Prebuilt Recurrent Reconstruction\\Network for Video Super-Resolution}

\author{Shuyun Wang*, Ming Yu*,~\IEEEmembership{Member,~IEEE,} Cuihong Xue,
        Yingchun Guo,~\IEEEmembership{Member,~IEEE,}  and Gang Yan% <-this % stops a space
\thanks{Shuyun Wang, Ming Yu, Yingchun Guo, and Gang Yan are with the School of Artiﬁcial Intelligence, Hebei University of Technology, Tianjin 300401, China.}
\thanks{Cuihong Xue is with Technical College for the Deaf, Tianjin University of Technology, Tianjin 300384, China.}
\thanks{Corresponding authors: Cuihong Xue (redxuech@tjut.edu.cn). *These authors contributed to the work equllly.}
\thanks{This work was supported by the National Natural Science Foundation of China (Grant No.61806071), the Open Projects Program of National Laboratory of Pattern Recognition (Grant No.201900043), the Sci-Tech Research Projects of Higher Education of Hebei Province, China (Grant No.QN2019207), and the Tianjin Sci-Tech development strategy research planning Projects (Grant No.18ZLZXZF00660).}}

\maketitle

\begin{abstract}
The video super-resolution (VSR) method based on the recurrent convolutional network has strong temporal modeling capability for video sequences. However, the temporal receptive field of different recurrent units in the unidirectional recurrent network is unbalanced. Earlier reconstruction frames receive less spatio-temporal information, resulting in fuzziness or artifacts. Although the bidirectional recurrent network can alleviate this problem, it requires more memory space and fails to perform many tasks with low latency requirements. To solve the above problems, we propose an end-to-end information prebuilt recurrent reconstruction network (IPRRN), consisting of an information prebuilt network (IPNet) and a recurrent reconstruction network (RRNet). By integrating sufficient information from the front of the video to build the hidden state needed for the initially recurrent unit to help restore the earlier frames, the information prebuilt network balances the input information difference at different time steps. In addition, we demonstrate an efficient recurrent reconstruction network, which outperforms the existing unidirectional recurrent schemes in all aspects. Many experiments have verified the effectiveness of the network we propose, which can effectively achieve better quantitative and qualitative evaluation performance compared to the existing state-of-the-art methods. 
\end{abstract}


% \fi
\IEEEpeerreviewmaketitle

\section{Introduction}
\IEEEPARstart{S}{UPER-RESOLUTION} task is designed to restore a low-resolution (LR) image or video sequence to the high-resolution (HR) counterpart by computing and filling the missing information. The super-resolution task involves two subtasks: single image super-resolution (SISR) \cite{dong2015image,shi2016real,ledig2017photo,niu2020single,wang2021unsupervised,menon2020pulse} and VSR \cite{yi2019progressive,tian2020tdan,kappeler2016video,li2020learning,yan2019frame,chu2020learning,isobe2020temporal}. In SISR, algorithms often rely on prior knowledge of natural images to restore HR details. In contrast, VSR, due to abundant temporal information contained in its video sequence, needs to restore both spatial information and temporal information. It has significant applicable value in monitoring devices, high-definition TV, satellite images, and video transmission. At the same time, when dealing with some other visual tasks \cite{li2020mucan,xiang2020zooming,kang2020deep,xu2021temporal}, the VSR algorithm may be used as a pre-algorithm to assist them in achieving better results. For that reason, how to extract more useful information from the time dimension and design an accurate, efficient and portable VSR algorithm is a problem that needs to be solved.

Recently, due to the strong time dependence of the recurrent convolutional network in modeling video, audio, and other sequence data processing, a large number of VSR methods based on recurrent architecture \cite{sajjadi2018frame,fuoli2019efficient,isobe2020revisiting,isobe2020video,huang2017video,li2019fast,chan2021basicvsr,zhu2019residual} have sprung up. Instead of using explicit optical flow \cite{wang2018learning} or motion compensation \cite{caballero2017real}, RLSP \cite{fuoli2019efficient} uses a recurrent convolution architecture to spread the temporal information in the form of the hidden state in high dimensional space. RRN \cite{isobe2020revisiting} adds residual blocks to the recurrent unit, which stabilizes the training of the recurrent convolution network. The increased network deep can simultaneously extract and fuse features more precisely, greatly improving the accuracy of super-resolution results. RSDN \cite{isobe2020video} divides the hidden state into two parts, structure and detail, then uses structure-detail block in the recurrent unit for processing. In addition, it also uses the hidden state adaptation module to screen the hidden state information needed to be used effectively. BasicVSR \cite{chan2021basicvsr} designed a bidirectional recurrent convolution architecture through experiments on each component, which significantly improves restoration quality of the model.

Although these VSR methods based on recurrent convolutional architecture enjoy good performance and low time consumption, the hidden state fed into the initial recurrent unit is all zero initialized, making the frames recovered from the early recurrent unit unable to obtain sufficient spatio-temporal information to reconstruction. Also, more useful information can often be obtained by the later recurrent units. This unbalanced information input method increases the difference in self-relative resolution of the reconstructed video, as is shown in Figure 1. In unidirectional recurrent convolutional \cite{sajjadi2018frame,fuoli2019efficient,isobe2020revisiting,isobe2020video,li2019fast,zhu2019residual} networks, there exist big resolution difference before and after the reconstructed video which exerts an influence on user experience and the video perception. Admittedly, the bidirectional recurrent convolutional network \cite{huang2017video,chan2021basicvsr} can alleviate this problem, the input sequence still needs to be reversed to propagate and complete the reconstruction. This propagation mechanism leads to high memory consumption and cannot output the reconstructed frames according to the time requirement, which greatly restricts its use in many application fields. This also increases complexity of computation, for the bidirectional recurrent convolutional network will unnecessarily reconstruct the later frame again.

Taking the issues all mentioned above into consideration, we thereby propose a portable Information Prebuilt Network. By extracting useful information from the earlier frames in the sequence, IPNet constructs the input hidden state to assist the recurrent unit to achieve better recovery of the earlier frames, which balances the overall recovery effect of the VSR. In addition, due to the better recovery effect of earlier frames, the subsequent recurrent unit can obtain more detailed spatio-temporal information, thus strengthening the recovery of the subsequent frames. Since the input of IPNet only requires part of the input frames in front of the video sequence, there is no need to do a reverse recurrent for all frames, which greatly reduces the calculation burden of the model. Moreover, it also is a portable component that can be applied to any unidirectional recurrent method to improve its reconstruction effect. 

\begin{figure}[!t]
\centering 
\includegraphics[width=3.2in]{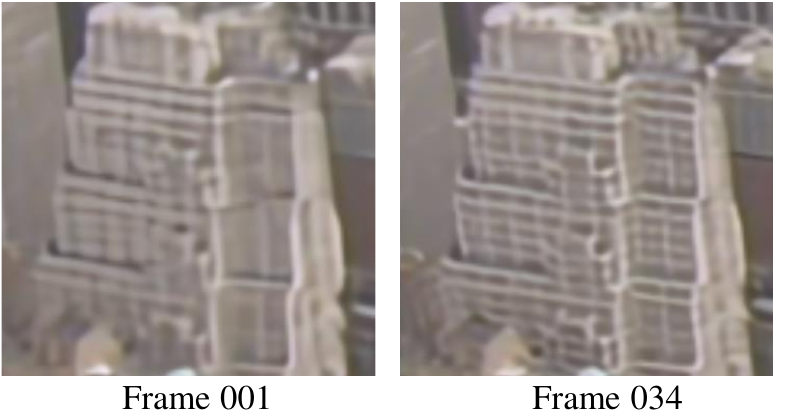}
\caption{The comparison of the restoration effect at different time steps on City of Vid4. RRNet is used here as a unidirectional recurrent network.}
\end{figure} 
 
Generally speaking, most of the previous VSR method with recurrent convolutional architecture use residual blocks \cite{he2016deep} as the main feature extraction and integration module. However, these models fail to make full use of the hierarchical features in temporal and spatial feature images. Inspired by RDN \cite{zhang2018residual}, we propose the RRNet that uses the residual dense block, to adapt the required feature information from the current or previous local features. In RRNet, the output of the propagation model is divided into two parts: the temporal dimension information and the spatial dimension information, which is different from the propagation of the previous recurrent method. To generate the temporal and spatial information, we use different operations and give them different tasks. The temporal information is more inclined to transfer the required time dimension information to the subsequent recurrent unit, while the spatial dimension information is needed to reconstruct the current frame. At the same time, the spatial information is also transmitted as auxiliary information to the subsequent recurrent unit to participate in the reconstruction of corresponding frames. The combination of the above IPNet and RRNet constitutes the proposed IPRRN.

In summary, the main contributions of this paper can be summarized as follows:
\begin{itemize}
\item We propose a novel information prebuilt network which balances the temporal receptive field at different time steps and reduces the resolution difference before and after the reconstructed video.

\item We introduce an efficient recurrent reconstruction network with significant recovery and propagation quality enhancements compared to existing unidirectional recurrent method.

\item Through a large number of experiments on IPRRN, we demonstrate that our method gives the state-of-the-art performance on most benchmark VSR datasets.
\end{itemize} 

The remainder of the paper is organized as follows. In section \uppercase\expandafter{\romannumeral2}, we describe related work. The third section introduces the details of our proposed VSR method. Section \uppercase\expandafter{\romannumeral4} introduces ablation experiments and comparison with the most advanced VSR methods, and we conclude our work in Section \uppercase\expandafter{\romannumeral5}.

\section{Related Work}
According to the different processing modes of a given frame sequence, existing VSR methods can be generally divided into the following two categories: sliding window method and recurrent method. The recurrent method can be further divided into unidirectional recurrent method and bidirectional recurrent method.

\subsection{Sliding window method}
Earlier sliding window methods predict the optical flow of input frames and used spatial warping for frame alignment. However, Inaccurate calculation of optical flow may lead to serious artifacts.  To improve the performance of VSR, other alignment modules are used in the following methods. Tian \emph{et al.} \cite{tian2020tdan} adopts deformable convolution to align inputs at the feature level. Jo \emph{et al.} \cite{jo2018deep} generates dynamic upsampling filters for implicit motion compensation. Wang \emph{et al.} \cite{wang2019edvr} designs a pyramid, cascading and deformable alignment module that uses deformable convolution in a multi-scale way to achieve better alignment. In addition, it also designs a temporal and spatial attention module to emphasize important features. Isobe \emph{et al.} \cite{isobe2020temporal} groups the temporal information and uses a mixture of 2D and 3D residual blocks for inter-group fusion. It cannot be ignored that, in these methods, the overlap between sliding windows leads to redundant calculations that limit the efficiency of VSR, and the temporal receptive field is limited by the size of the sliding window, which lacks long-term modeling capability. 

\subsection{Unidirectional recurrent method}
Recently, since the recurrent convolution network is able to calculate strong time consistency results, more and more methods employ it to deal with VSR tasks. The unidirectional recurrent method accesses video frames sequentially to complete reconstruction. Sajjadi \emph{et al.} \cite{sajjadi2018frame} puts the explicit optical flow into the recurrent unit, which reduces the extra calculation caused by repeated calculations of the sliding window. Fuoli \emph{et al.} \cite{fuoli2019efficient} introduces the high-dimensional hidden state into the recurrent unit and propagates the spatio-temporal information in the feature space. Isobe \emph{et al.} \cite{isobe2020revisiting} adds the residual block into the recurrent unit architecture to stabilize the training and boost the super-resolution performance. Yan \emph{et al.} \cite{yan2019frame} combines the local network and context network to get better reconstruction outputs. Isobe \emph{et al.} \cite{isobe2020video} divides the input into structure and detail components, then uses several structure-detail blocks for processing. Although these unidirectional recurrent methods have lower reconstruction time and better recovery effect, the temporal receptive field of the reconstruction units at different time steps is unbalanced, leading to the inconsistency of the relative resolution before and after the reconstruction frame sequence.

\subsection{Bidirectional recurrent method}
The hidden state of the bidirectional recurrent method propagates from forward and backward, which balances the temporal receptive field at different time steps. Huang \emph{et al.} \cite{huang2017video} first proposed a bidirectional recurrent convolutional architecture to deal with the VSR task. Chan \emph{et al.} \cite{chan2021basicvsr} designed a bidirectional recurrent method with embedded optical flow through the experiments of each component to enhance the recovery quality. Chan \emph{et al.} \cite{chan2021basicvsr} added the information refill mechanism and coupled propagation mechanism to promote information aggregation, further strengthening the recovery effect. Whereas these methods take the entire sequence as input for modeling and need to compute the sequence backward again for reconstruction, which requires high memory consumption, thus the methods are difficult to deploy on resource-constrained devices  and are not suitable for many low-latency application scenarios.

To tackle this problem, we thereby propose IPNet, a brand new network that has never been used before. IPNet uses partial frames in the front part of the frame sequence to calculate the initial hidden state, which stabilizes the reconstruction effect of the earlier frames. By enlarging the temporal receptive field of the earlier frames, the reconstruction resolution differences at different time steps can be greatly reduced. Combined with RRNet, a unidirectional recurrent network using feature reuse for efficient reconstruction, IPRRN can reconstruct videos that are more refined and smoother.

% or


\begin{figure}[!t] 
\centering
\includegraphics[width=3.2in]{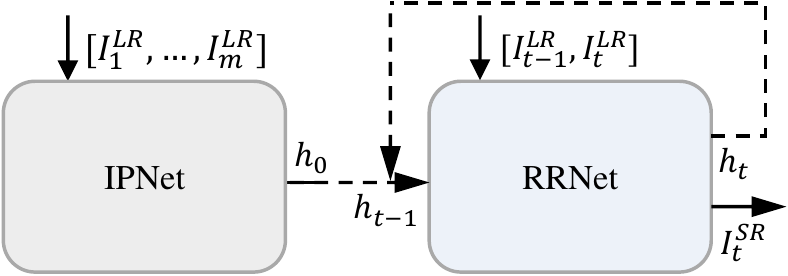}
\caption{The structure of IPRRN, which includes a IPNet and a RRNet. The IPNet generates the initial hidden state $h_0$ through the processing of the early m frames. The RRNet is an efficient unidirectional recurrent network, which recovers LR frame to corresponding SR frame by inputting two adjacent frames and the hidden state.}
\end{figure}

\begin{algorithm}[t]
\caption{Training process of IPRRN}
\begin{algorithmic}[1]
\REQUIRE ~~\\ %算法的输入参数：Input
Frame sequence of N LR frames $I^{LR}$;
\ENSURE ~~\\ %算法的输出：Output
Frame sequence of N super-resolution frames $I^{SR}$;
\FOR{max number of epochs}
\STATE Put the early m frames $[I^{LR}_1,...,I^{LR}_m]$ into the IPNet, then generates the initial hidden state $h_0$;
\FOR{$t=1$; $t<=N$; $t++$}
\IF{$t==1$} 
\STATE Let $I_{1}^{LR}$ replace $I_{t-1}^{LR}$;
\ENDIF 
\STATE Put the $I_{t}^{LR}$, $I_{t-1}^{LR}$, $h_{t-1}$ into the RRNet, then output the hidden state $h_t$ and the super-resolution frame $I_{t}^{SR}$;
\ENDFOR
\STATE Minimize loss L1$(I^{LR},I^{SR})$;
\ENDFOR
\end{algorithmic}
\end{algorithm}


\section{Methodology}
This section will introduce our proposed IPRRN. First, we introduce the overall architecture of our method and then give the details of each module.

\subsection{Overview}
Given a frame sequence $I^{L R} = \left\{I_{n}^{L R}\in \mathbb{R}^{H \times W \times C}\right\}_{n=1}^{N}$ contains N LR frames, our IPRRN is designed to reconstruct the corresponding super-resolution frame sequence $I^{S R} = \left\{I_{n}^{S R}\in \mathbb{R}^{sH \times sW \times C}\right\}_{n=1}^{N}$ with an upscale factor $s$, while making it closer to the ground truth frame sequence $I^{H R} = \left\{I_{n}^{H R}\in \mathbb{R}^{sH \times sW \times C}\right\}_{n=1}^{N}$. Figure 2 shows the architecture of our IPRRN, which consists of two major networks: IPNet and RRNet. IPNet receives early m (1$\leq$ m $\leq$N) frames as input and generates the initial hidden state $h_0$ required by the first RRNet unit, $h_0$ is generated as follows:
\begin{equation}
h_0 = f_{IPNet}(I^{LR}_1,...,I^{LR}_m)
\end{equation}Where $f_{IPNet}(\cdot)$ represents the IPNet. It is worth noting that IPNet performs only once during the reconstruction of the entire sequence. It is scheduled to operate before the RRNet to estimate the hidden state $h_0$ required for the first recurrent unit. Then we use RRNet to reconstruct the frame sequence in a unidirectional way. At time step $t$, the input of the recurrent unit is divided into two parts: the hidden state $h_{t-1}$ computed by the previous recurrent unit and the reference frame $I_t^{LR}$ with the neighboring frame $I_{t-1}^{LR}$. These inputs are first concatenated on the channel dimension, then 2D convolution, ReLU activation function, and residual dense blocks (RDBs) are used for deep feature extraction and integration of features. Finally, the hidden state $h_t$ and the predicted super-resolution results $I_t^{SR}$ are generated by the reconstruction module:
\begin{equation}
h_t,I_t^{SR} = f_{RRNet}(h_{t-1},I_{t-1}^{LR},I_t^{LR})
\end{equation}Where $f_{RRNet}(\cdot)$ represents the RRNet. Algorithm 1 shows the entire training process of our IPRRN.

\subsection{Information Prebuilt Network}
Figure 3 shows the structure of IPNet. Compared with other unidirectional recurrent methods, which use the all-zero assignment method to generate the initial hidden state, IPNet generates the initial hidden state through the processing of the early m LR frames $[I^{LR}_1,...,I^{LR}_m]$. This balances the temporal receptive field at different time steps, thus reducing the relative visual gap in the reconstructed video at different moments. Our IPNet includes three modules: shallow feature extraction and filter module, deep feature extraction module, and propagation module.

\begin{figure*}[ht] 
\centering
\includegraphics[width=7.1in]{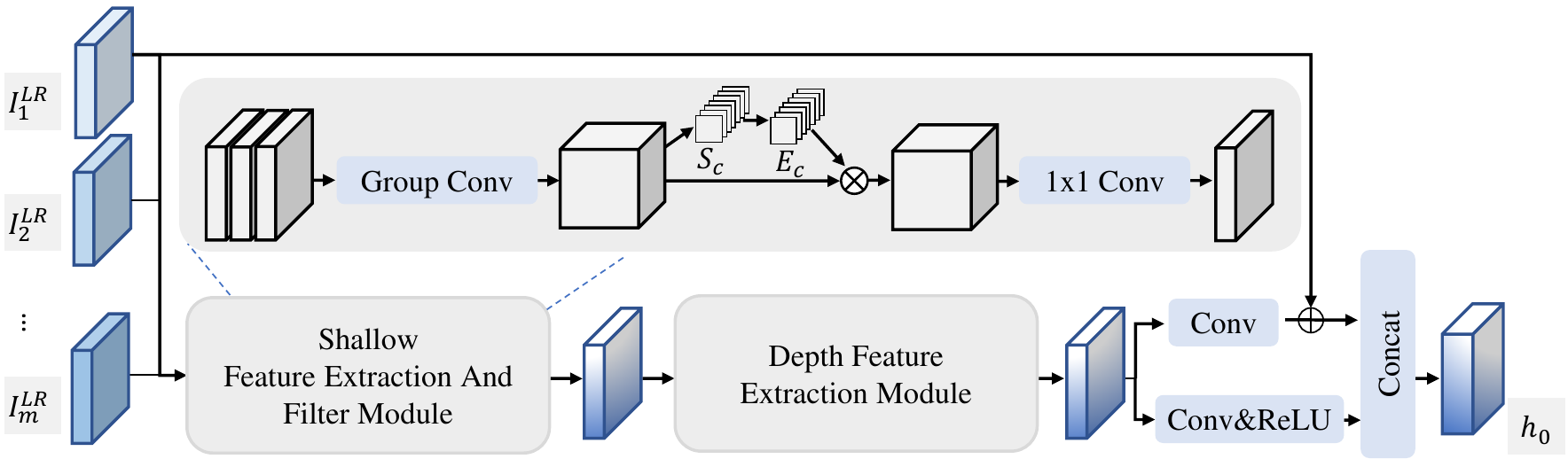}
\caption{Structure of proposed IPNet. First, early m frames were fed into a group convolution to extract shallow features, then the SE block and a 1$\times$1 convolution were used for feature screening, and then input the features into the deep feature extraction module. Finally, the propagation module was used to normalized the deep features into hidden state space.}
\end{figure*}  

\subsubsection{Shallow feature extraction and filter module}
We use a group convolution to extract the shallow feature $F_{sf}$ from the input m frames. Since the information redundancy, which makes the reconstruction of earlier frames or the subsequent propagation inefficient, exists in the shallow feature, we feed the shallow feature into a filter module consisting of a channel attention \cite{zhang2018image} and feature dimension reduction to eliminate the redundancy. Here, we use the Squeeze and Congestion (SE) block from SENet \cite{hu2018squeeze} for the channel attention, as is shown in Figure 3. SE block first performs a Squeeze operation on the feature map obtained by convolution and uses global average pooling to encode all features in the space as global features $S_{c} \in \mathbb{R}^{1 \times 1 \times C}$:
\begin{equation}
S_{c}=\frac{1}{H \times W} \sum_{i=1}^{H} \sum_{j=1}^{W} F_{sf}(i, j)
\end{equation}
Then excitation the global features and compute the relationships between different channels to get the weight $E_c$:
\begin{equation}
E_{c}=\sigma(W_{2} *\operatorname{ReLU}(W_{1} * S_{c}))
\end{equation}
Where $W_1$ and $W_2$ are the weight matrix, '* 'is the matrix multiplication operation, $\sigma(\cdot)$ and ReLU$(\cdot)$ are the sigmoid activation function and ReLU activation function. Finally, the different channels are multiplied by their corresponding weights and use a convolution operation with 1x1 kernel size for feature dimension reduction to get the filtered shallow feature $F_{ff}$:
\begin{equation}
F_{ff}=\operatorname{Conv}_{1}\left(E_{c}\odot F_{sf}\right)
\end{equation}
Where $\odot$ denotes channel-wise multiplication, $\operatorname{Conv_1}(\cdot)$ represents the convolution operation with 1x1 kernel size. Channel attention and feature dimension reduction enable the network to selectively enhance useful features in the channel dimension and suppress repeated features, so that our model can obtain the required information for the reconstruction of earlier frames and improve the computing efficiency of the IPNet.

\subsubsection{Deep feature extraction module}
The deep feature extraction module is composed of several stacked residual blocks. We take the filtered shallow feature as input to extract the deep feature information $F_{df}$:
\begin{equation}
  F_{df} = f_{DE}(F_{ff})
\end{equation}
Then we put the $F_{df}$ into the propagation module to calculate the initial hidden state.

\subsubsection{Propagation module}
As is shown in Figure 3, to make the RRNet use the extracted deep feature better, we use the proposed propagation module to map the deep feature $F_{df}$ into the hidden state dimension space, so that the subsequent recurrent unit can directly understand it and effectively use it:
\begin{equation}
h_0 = f_{PM}(F_{df})
\end{equation}
Where $f_{PM}$(·) is the propagation module. Since the propagation module is a part of propagation and reconstruction module in RRNet, these two modules will be following illustrated together in the propagation and the reconstruction module.

\subsection{Recurrent Reconstruction Network}
The proposed RRNet is based on the unidirectional recurrent convolution architecture. It is worth noting that our recurrent unit corresponds to the order of the reference frame. Specifically, the reference frame corresponding to the $t$th recurrent unit is $I_t^{LR}$, and the corresponding super-resolution frame is $I_t^{HR}$. Figure 4 shows the architecture of the RRNet. It consists of two modules: feature extraction with integration module and propagation and reconstruction module.

\subsubsection{Feature extraction with integration module}  
At time step $t$, the input of RRNet consists of two parts: one is the hidden state $h_{t-1}$ calculated by the recurrent unit at time step $t-1$, the other is the reference frame $I_t^{LR}$ and the neighboring frame $I_{t-1}^{LR}$. These inputs are concatenated on the channel dimension, then extracted by a 3x3 convolution with ReLU activation for shallow feature extraction, as is shown in the following formula:
\begin{equation}
F_{t}=\operatorname{ReLU}\left(\operatorname{Conv}_{3}\left(\left[I_{t}^{L R}, I_{t-1}^{L R}, h_{t-1}\right]\right)\right)
\end{equation}
Where $F_t$ represents the shallow feature in RRNet. Then, we feed $F_t$ into the deep feature extraction module consisting of multiple RDBs. Most previous methods use residual block for feature extraction, whether the local sliding window method or recurrent method. This leads to these models cannot make full use of the hierarchical feature they generated. To solve this problem, we proposed RRNet by using RDB for feature extraction, this is more friendly for low-level semantic feature. RDB can get the features it needs from the current or previous local features adaptively to complete more refined feature extraction and integration. In our approach, each residual dense block consists of three convolutions of 3x3 with ReLU activation function and a 1x1 convolution series. The deep feature $D_t$ is calculated through RDBs, as is shown in the following formula:
\begin{equation}
D_t = RDBs(F_t)
\end{equation}
 
\begin{figure}[!t]
\centering
\includegraphics[width=3.4in]{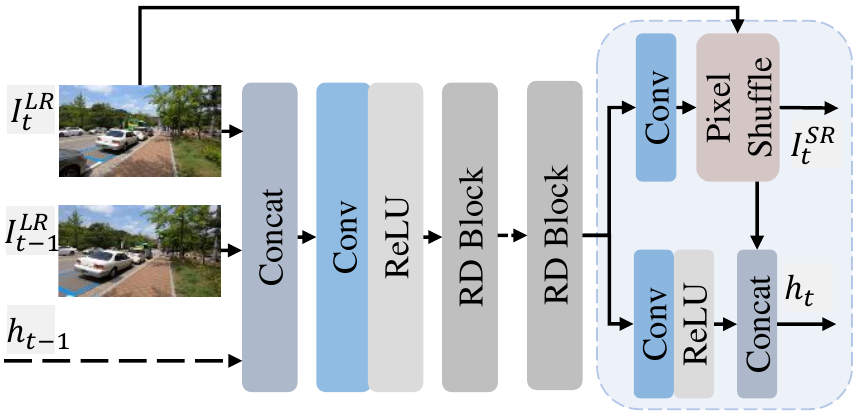}
\caption{Structure of proposed RRNet. RRNet is based on the unidirectional recurrent convolutional network, and it uses RDBs to extract and fuse features sufficiently to reconstruct the reference frame.}
\end{figure} 
 
\subsubsection{Propagation and reconstruction module}
The detail of propagation and reconstruction module is shown in the right box in Figure 4. Since the VSR task needs not only to improve the spatial resolution but also to model in the temporal dimension, the effective use of spatio-temporal information becomes an important factor affecting the quality of reconstruction.In RRNet, the hidden state summarizes the information from the previous frame and current frame. To take full advantage of temporal and spatial information, the generation of hidden state in our RRNet is divided into two parts. One is the temporal information, which integrates the past temporal information in the form of a high-dimensional feature. The other is spatial information, which pays more attention to intra-frame details to better complete the reconstruction. The temporal information is generated by convolution with ReLU activation, while another convolution generates the spatial information. The two parts are concatenated on the channel dimension as the output hidden state. The above is the specific structure of the propagation module in propagation and reconstruction module, which is also the structure of the propagation module in IPNet. To recover the super-resolution frame, we convert the spatial information into the residual frame in the super-resolution space $K_t^{SR}$ utilizing pixel shuffle operation. Meanwhile, $I_t^{LR}$ performs bicubic interpolation and then adds to $K_t^{SR}$ to output the super-resolution frame $I_t^{HR}$. IPNet needs to pass the hidden state backwards without calculating the reconstruction frame, so the propagation module only generates the initial hidden state.

\section{Experiments}
In this section, we will introduce the datasets for model training and testing. Then, we verify the effectiveness of different modules in our proposed network. Finally, our results are evaluated qualitatively and quantitatively with other methods.
\subsection{Dataset}
To make a fair comparison with other methods, in this work, we consider two widely-used datasets Vimeo-90K \cite{xue2019video} and REDS \cite{nah2019ntire}. For REDS \cite{nah2019ntire}, it contains 300 video sequences, in which 240 for training, 30 for validation, and 30 for testing. Each sequence contains 100 frames with a resolution of 720 $\times$ 1280. Following \cite{chan2021basicvsr}, we use the REDS4 \cite{nah2019ntire} dataset as our test set and the rest 266 sequences from the training and validation set as the training set. Same with previous works, we apply the Bicubic degradation on REDS4 \cite{nah2019ntire}, and the results are calculated on RGB-channel. For Vimeo-90K \cite{xue2019video}, it consists 89,800 video clips of a rich number of scenes and actions. We follow the standard training and testing dataset splits, and the training set contains 64,612 videos. Each video has seven frames with the resolution of 448 $\times$ 256. To generate the LR video, we intercept the original video frame into 256 $\times$ 256 HR patch, then the Gaussian kernel of $\sigma=1.6$ was used for filtering, and the scale factor of 4$\times$ was further sampled to obtain the LR patch with size 64 $\times$ 64. After training the model on Vimeo-90K \cite{xue2019video}, we tested it on several popular benchmark datasets: Vid4 \cite{liu2011bayesian}, Vimeo-90K-T \cite{xue2019video} and SPMCS \cite{tao2017detail}. We used peak signal-to-noise ratio (PSNR) and structural similarity index (SSIM) \cite{wang2004image} to calculate the luminance (Y) channel in the transformed YCbCr space and evaluated the proposed model by comparing its performance with the latest methods.

\subsection{Implementation details}
We use the L1 loss as loss function, which is defined as $L=\Vert I_t^{HR}-I_t^{SR}\Vert$. The IPRRN used for the comparison (called Ours) contains 10 RDBs, and temporal information and spatial information in the hidden layer have 128 and 48 channels respectively. Unless noted otherwise, the IPNet takes seven consecutive early frames (i.e., m=7) as the input. Each mini-batch consists of 8 patches, and we use the optimizer Adam \cite{kingma2014adam} to update them, where $\beta_1$=0.9,$\beta_2$=0.999. The initial learning rate is set to 1e-4, and is later down-scaled by a factor of 0.1 every 60 epoch. All experiments were performed on a server with PyTorch 1.7 and Nvidia Rtx 3090 GPU.

\subsection{Ablation Studies}
To further verify our model, we performed ablation studies, for example, replacing or removing partial components of our framework. In this section, we present the ablation results on the Vid4 \cite{liu2011bayesian} test set to prove that our design is correct.

\begin{figure*}[!t] 
\centering
\includegraphics[width=7.1in]{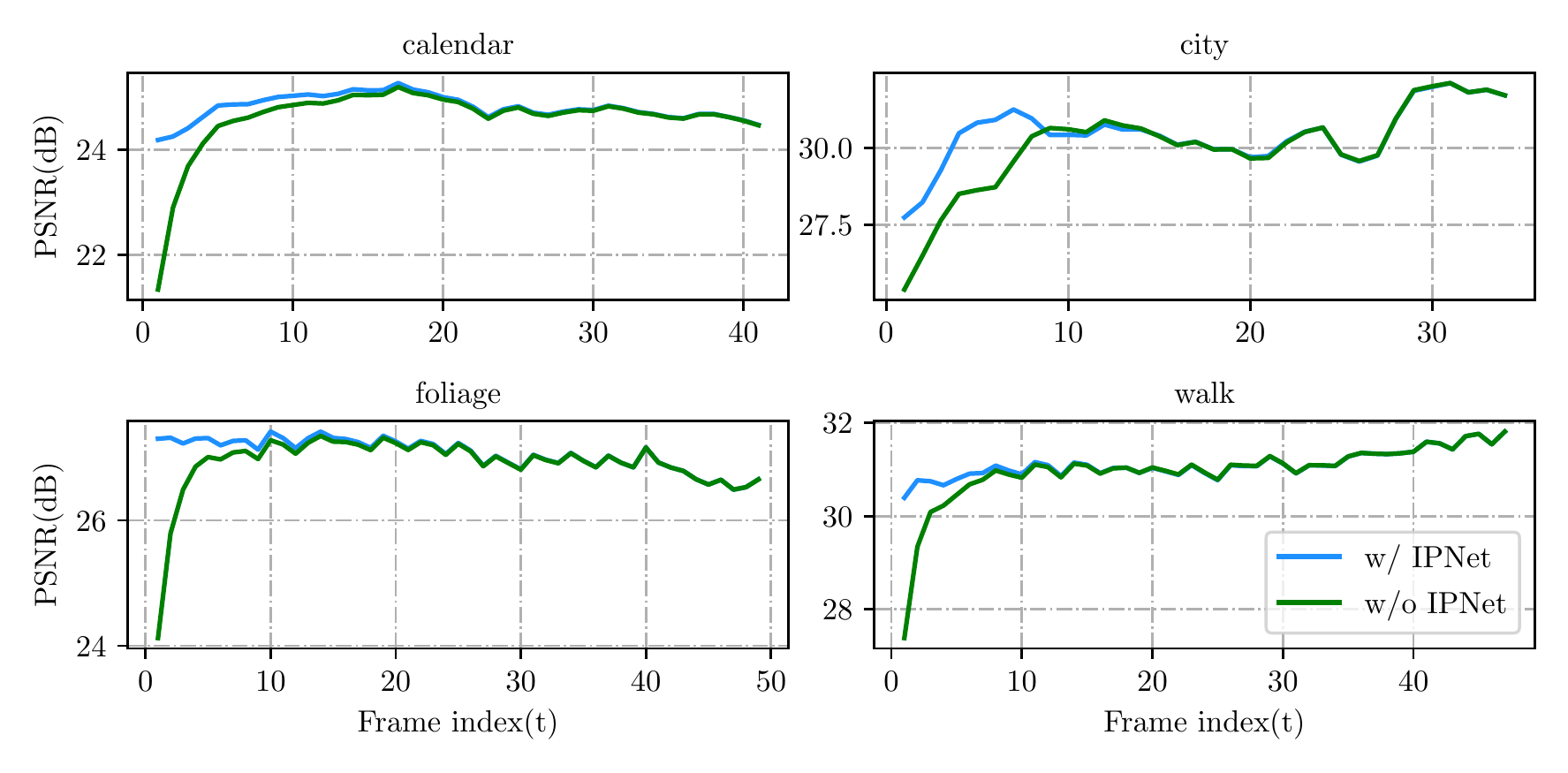}
\caption{RRNet compared to IPNet+RRNet on Vid4. The PSNR of the initial frame is relatively low when IPNet does not exist. After adding IPNet, the performance of earlier frames has improved dramatically, and this improvement will be continued during the length of the temporal receptive field.}
\end{figure*}

\begin{table}[t]
\caption{Study on relative resolution gap in super-resolution video of Vid4, where gap is defined as the difference between the maximum PSNR and the minimum PSNR in super-resolution video}
\centering  % 表居
\begin{tabular}{l|c|c|c|c|c|c}
\hline
\multirow{2}{*}{ \text {Scene}} &
\multicolumn{3}{c|}{ \text {w/o IPNet}} &
\multicolumn{3}{c}{ \text {w/ IPNet}} \\
\cline{2-7}  %  \cline用于画横线 \cline{i-j}表示从第i列画到第j列
&\makebox[0.04\textwidth][c]{ \begin{tabular}{c}  \text {Min} \\  \text {PSNR} \end{tabular}}
&\makebox[0.04\textwidth][c]{\begin{tabular}{c} \text { Max} \\  \text {PSNR} \end{tabular}} 
& \makebox[0.04\textwidth][c]{ \begin{tabular}{c}  \text {Gap} \\  \text {PSNR} \end{tabular}} 
&\makebox[0.04\textwidth][c]{\begin{tabular}{c} \text {Min} \\ \text {PSNR} \end{tabular}}
&\makebox[0.04\textwidth][c]{\begin{tabular}{c} \text { Max }\\  \text {PSNR} \end{tabular}} 
& \makebox[0.05\textwidth][c]{ \begin{tabular}{c}  \text {Gap} \\  \text {PSNR(dB)} \end{tabular}}\\
\hline \hline
\text {Calendar} & 21.35 & 25.18 &3.83 & 24.18 & 25.26 & 1.08 \\
\text {City} & 25.37 & 32.14 & 6.77 & 27.73 & 32.14 & 4.41 \\
\text {Foliage} & 24.12 & 27.35 & 3.23 & 26.49 & 27.42 & 0.93 \\
\text {Walk} & 27.38 & 31.81 & 4.43 & 30.39 & 31.81 & 1.42 \\
\hline
\text {Average} & 24.56 & 29.12 & 4.56 & 27.20 & 29.16 & 1.96 \\
\hline
\end{tabular}
\end{table}

\subsubsection{Information prebuilt network}
Since the IPNet has not been mentioned in the previous methods, we specially designed the ablation study for this component to verify the effectiveness of IPNet. Table \uppercase\expandafter{\romannumeral1} shows the relative resolution gap in super-resolution video, where the gap is defined as the differences between the maximum PSNR and the minimum PSNR in super-resolution video. It can be observed that the average PSNR gap of the super-resolution video without IPNet is 4.56dB, and the gap reduces to 1.96dB when IPNet is available, a drop of 57.02\%. These indicate that IPNet greatly enhances the consistency of the reconstruction video. Figure 5 makes comparison of the PSNR values of output images with and without IPNet. The PSNR of the initial frame is relatively low when IPNet does not exist. After IPNet is added, the performance of earlier frames has improved dramatically, and this improvement will continue during the length of the temporal receptive field. It is worth noting that the value of the length of the temporal receptive field is bigger than the number of input frames in IPNet. This is because better reconstruction can fine the hidden state, thus providing more useful spatio-temporal information for the recovery of subsequent frames. As the temporal receptive field of the reference frame becomes saturated, this gain disappears. In addition, Figure 6 shows the PSNR value of the first frame of our method compared with other unidirectional recurrent methods on Vid4 \cite{liu2011bayesian}. As we can see, IPNet greatly exceeds the state-of-the-art unidirectional recurrent methods in all four scenarios, which improves the overall recovery level and reduces the relative resolution differences in super-resolution video. Moreover, IPNet is also portable, that is, it can be applied to other unidirectional recurrent methods to improve their recovery effect.

\begin{figure}[tp]
\centering
\includegraphics[width=3.4in]{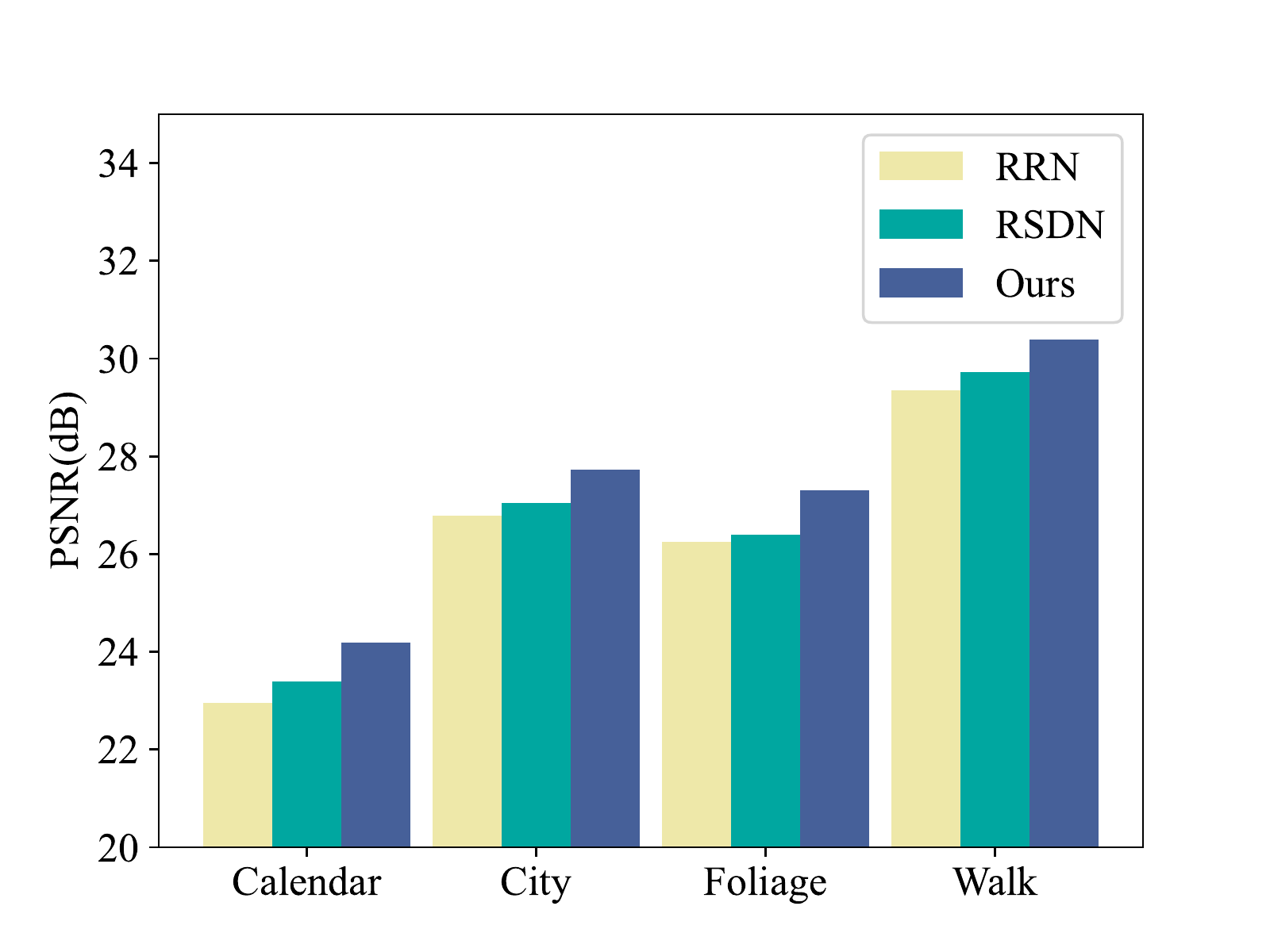}
\caption{Quantitative evaluation with other unidirectional recurrent methods on the first frame on Vid4}
\end{figure}

\begin{table*}[htp]
\caption{Quantitative comparison on Vid4 for 4$\times$ VSR. Red text
indicates the best and blue text indicates the second best performance (PSNR/SSIM). FLOPs is computed on an LR size of 180$\times$320.}
\centering  % 表居中 
\begin{tabular}{l|c|c|cccc|c}
\hline \text { Method } & \text { Params(M) } & \text { FLOPs(T) } & \text { Calendar } & \text { City } & \text { Foliage } & \text { Walk } & \text { Average } \\
\hline \text { Bicubic } & - & - & 20.39 / 0.5720 & 25.16 / 0.6028 & 23.47 / 0.5666 & 26.10 / 0.7974 & 23.78 / 0.6347 \\
\text { FRVSR  \cite{sajjadi2018frame}} & 5.1& 0.34  & 23.46 / 0.7854 & 27.70 / 0.8099 & 25.96 / 0.7560 & 29.69 / 0.8990 & 26.70 / 0.8126 \\
\text { DUF \cite{jo2018deep}} & 5.8& 2.29  & 23.85 / 0.8052 & 27.97 / 0.8253 & 26.22 / 0.7646  & 30.47 / 0.9118  & 27.13 / 0.8267 \\
\text { RBPN \cite{haris2019recurrent}}  & 12.2& 8.52  & 24.33 / 0.8244 & 28.28 / 0.8413  & 26.46 / 0.7753 & 30.58 / 0.9130 & 27.41 / 0.8385 \\
\text { PFNL \cite{yi2019progressive}} & 3.0 & 0.94 & 24.37 / 0.8246 & 28.09 / 0.8385 & 26.51 / 0.7768 & 30.65 / 0.9135 & 27.40 / 0.8384 \\
\text { EDVR \cite{wang2019edvr}} & 3.3& 2.95  & - & - & - & - & 27.85 / 0.8503 \\
\text { TGA \cite{isobe2020temporal}} & 7.1& 0.70  & 24.47 / 0.8286 & 28.37 / 0.8419 & 26.59 / 0.7793 & 30.96 / 0.9181 & 27.59 / 0.8419 \\
\text { RLSP \cite{song2021multi}} & 4.2& 0.31  & 24.60 / 0.8355 & 28.14 / 0.8453 & 26.75 / 0.7925 & 30.88 / 0.9192 & 27.60 / 0.8476 \\
\text { RSDN \cite{isobe2020video}}  & 6.2& 0.36  & 24.60 / 0.8345 & \textcolor{blue}{29.20 / 0.8527} & 26.84 / 0.7931 & 31.04 / 0.9210 & 27.92 / 0.8505 \\
\text { RRN \cite{isobe2020revisiting}}  & 3.4& 0.19  & 24.55 / 0.8342 & 28.55 / 0.8478 & 26.94 / 0.7983 & 30.75 / 0.9171 & 27.69 / 0.8488 \\
\text { BasicVSR \cite{chan2021basicvsr}} & 6.3& 0.33 & 24.68 / 0.8413 & 28.73 / 0.8564 & 26.90 / 0.7972 & \textcolor{blue}{31.53 / 0.9264} & 27.96 / 0.8553 \\
\text { IconVSR \cite{chan2021basicvsr}} & 8.7& 0.51 & \textcolor{red}{24.90 / 0.8455}  & 28.45 / 0.8545 & \textcolor{red}{ 27.07 / 0.7999} & \textcolor{red}{31.61 / 0.9268}  & \textcolor{blue}{28.01 / 0.8567} \\
\hline \text { IPRRN(ours) }  & 6.1 & 0.24 & \textcolor{blue}{24.77 / 0.8417} & \textcolor{red}{30.52 / 0.8700} & \textcolor{blue}{27.05 / 0.8011} & 31.11 / 0.9223 & \textcolor{red}{28.36 / 0.8588} \\
\hline
\end{tabular}
\end{table*}

\begin{table*}[t]
\caption{Comparative study of RRNet with other unidirectional recurrent convolutional networks on Vid4, runtime is computed as the average of 100 LRs of size 180$\times$320}
\centering  % 表居中
\begin{tabular}{l|ccc|ccc}
\hline
\multirow{2}{*}{\text { Method }} & \multicolumn{3}{c|}{w/o IPNet} & \multicolumn{3}{c}{w/ IPNet} \\
\cline{2-7}& \text { Params(M) } & \text { Time(ms) } & \text { PSNR(dB)/SSIM } & \text { Params(M) } & \text { Time(ms) } & \text { PSNR(dB)/SSIM }\\
\hline \text { RLSP \cite{song2021multi}} & 5.55 & 49 & 27.48 / 0.8388 & - & - &-\\
\hline \text { RRN \cite{isobe2020revisiting}} & 3.40 & 45 & 27.69 / 0.8488 & 5.35 & 46 & 27.99 / 0.8539 \\
\hline \text { RSDN \cite{isobe2020video}} & 6.18 & 94 & 27.92 / 0.8505 & 9.48 & 96 &  28.16 / 0.8551 \\
\hline \text { RRNet } & 4.14 & 56 & 27.92 / 0.8523 & 6.10 & 57 & 28.36 / 0.8588  \\
\hline
\end{tabular}
\end{table*}

\begin{table}[htp]
\caption{Study on the different number of input frames in IPNet on Vid4}
\centering  % 表居中
\begin{tabular}{c|c}
\hline \text { IPNet Inputs } & \text { PSNR(dB)/SSIM } \\
\hline 0 & 27.92 / 0.8523 \\
\hline 3 & 28.07 / 0.8532 \\
\hline 5 & 28.21 / 0.8560 \\
\hline 7 & 28.36 / 0.8588 \\
\hline
\end{tabular}
\end{table}

\begin{table}[tp]
\caption{Verify the validity of various components in the framework on Vid4}
\centering  % 表居中
\begin{tabular}
{ccc|c|c}
\hline 
RRNet &\begin{tabular}{c} IPNet w/o \\SE Block \end{tabular} &IPNet &PSNR(dB) &Params(M) \\
\hline \checkmark & & & 27.92 & 4.14 \\
\hline \checkmark & \checkmark & & 28.05 & 6.07 \\
\hline \checkmark & & \checkmark & 28.36 & 6.10\\
\hline
\end{tabular}
\end{table}

\begin{table}[htp]
\caption{Study on the number of rdbs in RRNet on Vid4}
\centering  % 表居中
\begin{tabular}
{c|c}
\hline \text { RDB number } & \text { PSNR(dB)/SSIM } \\
\hline 5 & 27.28 / 0.8387 \\
\hline 7 & 27.75 / 0.8485 \\
\hline 10 & 27.92 / 0.8523 \\
\hline
\end{tabular}
\end{table}

\subsubsection{Portability of IPNet}
We transplant IPNet to other unidirectional recurrent methods, RRN \cite{isobe2020revisiting} and RSDN\cite{isobe2020video}, to verify the validity of our module. Due to the differences exist in feature processing and feature spaces, we fine-tune the network after adding IPNet to make the high-dimensional space of initial hidden state calculated by IPNet and the feature space processed by recurrent network could be better matched. As is shown in Table \uppercase\expandafter{\romannumeral3}, we can find that RRN \cite{isobe2020revisiting} and RSDN \cite{isobe2020video} with IPNet increase by 0.30dB and 0.24dB respectively, this proves that our IPNet can be transplanted into any unidirectional recurrent networks which has imbalance temporal receptive field to improve their performance.
    
\subsubsection{The number of input frames of IPNet}
This section further discusses the influence of different numbers of input frames in IPNet on reconstruction performance. It can be seen from Table \uppercase\expandafter{\romannumeral4}, the reconstruction performance improves significantly as the number of input frames increases. When the number equals zero, the output of IPNet, initial hidden state, will be equal to all zero-initialized. At this time is 27.92dB. When IPNet only receives information from the initial three adjacent frames, the PSNR value is 28.07dB, 0.15dB higher than the zero-frame input. As the number of input frames increases from three to seven, the reconstruction turns out to be more accurate. This phenomenon is consistent with our theoretical reasoning, because more input frames contain more recovery information, which is very helpful for recovering the structure and details of the reference frame. Since there are only seven frames available in each video of Vimeo-90K \cite{xue2019video} dataset, the maximum input frame of this experiment is only seven. In fact, as the number of input frames increases, the performance of the reconstruction correspondingly improves, because more valuable spatial and temporal information can be obtained.

\begin{table*}[ht]
\caption{Quantitative comparison on REDS4 \cite{nah2019ntire}, SPMCS \cite{tao2017detail} and Vimeo-90K-T \cite{xue2019video} for 4$\times$ VSR. All the results are calculated on Y-channel except REDS4 \cite{nah2019ntire} (RGB-channel). Red textindicates the best and blue text indicates the second best performance (PSNR/SSIM). Blanked entries correspond to results unable to be reported. $'\dagger'$ means part of the values are calculated using provided models.}
\centering  % 表居中 
\begin{tabular}{l||c|c|c||c|c|c}
\hline \text { Method }& \makebox[0.08\textwidth][c]{\text { \#Frame}} & \text { Runtime(ms) }& \makebox[0.09\textwidth][c]{\text { FPS }} & \text { REDS4 \cite{nah2019ntire}} &\text { SPMCS \cite{tao2017detail}} & \text { Vimeo-90K-T \cite{xue2019video}}\\
\hline 
\text { Bicubic }&1 &-&-& 26.14 / 0.7292  & 23.29 / 0.6385 & 31.30 / 0.8687 \\
\text { DUF \cite{jo2018deep} } &7& 974 & 1.0 & 28.66 / 0.8262 &  29.63 / 0.8719 & 36.87 / 0.9447 \\
\text { RBPN \cite{haris2019recurrent}} &7 & 1507 & 0.7 & 30.09 / 0.8590& 29.54 / 0.8704 & 37.20 / 0.9458 \\
\text { PFNL \cite{yi2019progressive}}  &7& 295 & 3.4 & 29.63 / 0.8502 & 30.02 / 0.8804 & - \\
\text { EDVR \cite{wang2019edvr}}  &7& 378 & 2.6 & 31.09 / 0.8800 & - &  \textcolor{blue}{37.81 / 0.9523}\\
\text { TGA \cite{isobe2020temporal}} &7 & 375 & 2.7 & -  & 30.31 / 0.8857 & 37.59 / 0.9516 \\
\text { FRVSR \cite{sajjadi2018frame}}  &uni& 137 & 7.3 & - & 28.16 / 0.8421 & 35.64 / 0.9319 \\
\text { RLSP \cite{song2021multi}} &uni& 49 & 20.4 & -   & 29.64 / 0.8791 & 36.49 / 0.9403 \\
\text { RSDN \cite{isobe2020video}}&uni & 94 & 10.6 & - & 30.18 / 0.8811&37.23 / 0.9471\\
\text { RRN \cite{isobe2020revisiting}} &uni& 45 & 22.2 & -  & 29.83 / 0.8824 & - \\
\text { BasicVSR$^\dagger$  \cite{chan2021basicvsr}}&bi & 63 & 15.9 & \textcolor{blue}{31.42 / 0.8909} & 30.24 / 0.8732  & 37.53 / 0.9498  \\
\text { IconVSR$^\dagger$  \cite{chan2021basicvsr}} &bi & 70 & 14.3 & \textcolor{red}{31.67 / 0.8948}& \textcolor{blue}{30.28 / 0.8755} & \textcolor{red}{37.84 / 0.9524}\\
\hline\text { IPRRN(ours) } &7+uni& 57 & 17.5 & 31.13 / 0.8821 & \textcolor{red}{30.52 / 0.8911} & 37.53 / 0.9507 \\
\hline
\end{tabular}
\end{table*} 

\begin{figure*}[!ht] 
\centering 
\includegraphics[width=7.1in]{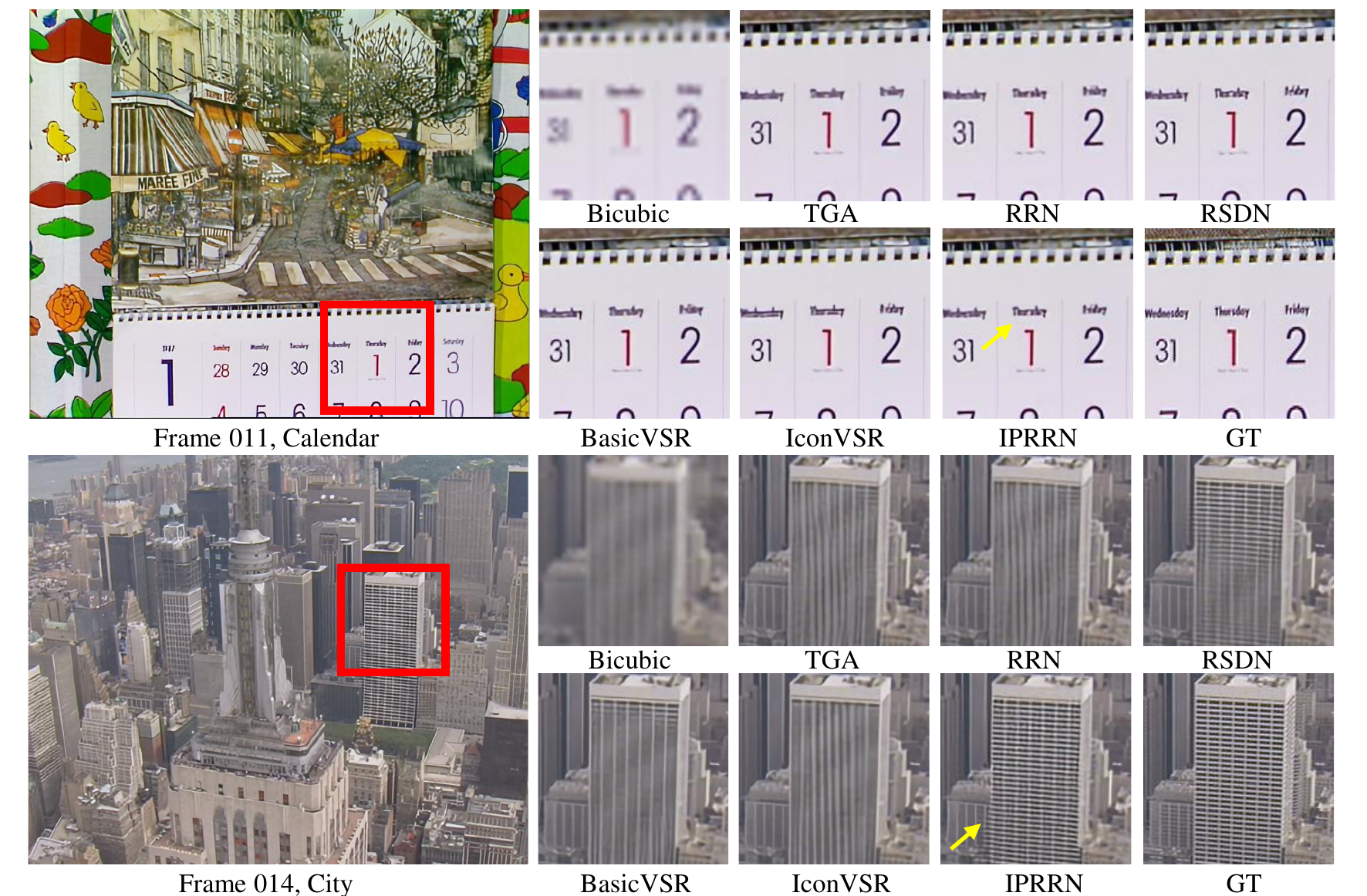}
\caption{Super-resolution qualitative comparison with sampling factor $\times$4 on Vid4. Zoom in for better visualization.}
\end{figure*}
          
\subsubsection{SE Block}
As is shown in Table \uppercase\expandafter{\romannumeral5}, we can observe that RRNet can achieve 27.92dB on Vid4 \cite{liu2011bayesian}, and it increased by 0.13dB after IPNet without SE block is added. With the support of SE Block, our network achieved the highest PSNR score of 28.36dB in the test, rising by 0.48dB and 0.31dB compared to the previous two schemes. This is because the convolution operation lacks the ability to constrain features, and features calculated by the IPNet without SE block will get too much redundant information, leading to inefficient propagation of hidden states. After adding the SE block, some channels are suppressed due to similar contents, while some channels are enhanced because they possess relatively independent and effective information. The enhancement and the suppression enable IPNet to pay more attention to the information it needs to refine the quality of the propagation content and strengthen the  efficiency and effectiveness of the recovery of earlier frames.

\subsubsection{Performance study of RRNet}
We compared the RRNet with other unidirectional recurrent methods from many aspects to show the advantages of RRNet. According to the Table \uppercase\expandafter{\romannumeral3}, compared with the RSDN \cite{isobe2020video} method, our model has a faster speed and smaller parameters when the recovery performance is approximately the same. Compared with RRN \cite{isobe2020revisiting} and RLSP \cite{fuoli2019efficient}, with only a small increase in parameters and computation time, we have achieved a significant improvement in the recovery effect. These enhancements are dependent on the reuse of features at different levels in RDBs to generate finer features and hidden states for reconstruction and propagation, reflecting that our RRNet is the most efficient approach among unidirectional recurrent methods.

\begin{figure*}[!ht] 
\centering
\includegraphics[width=7.1in]{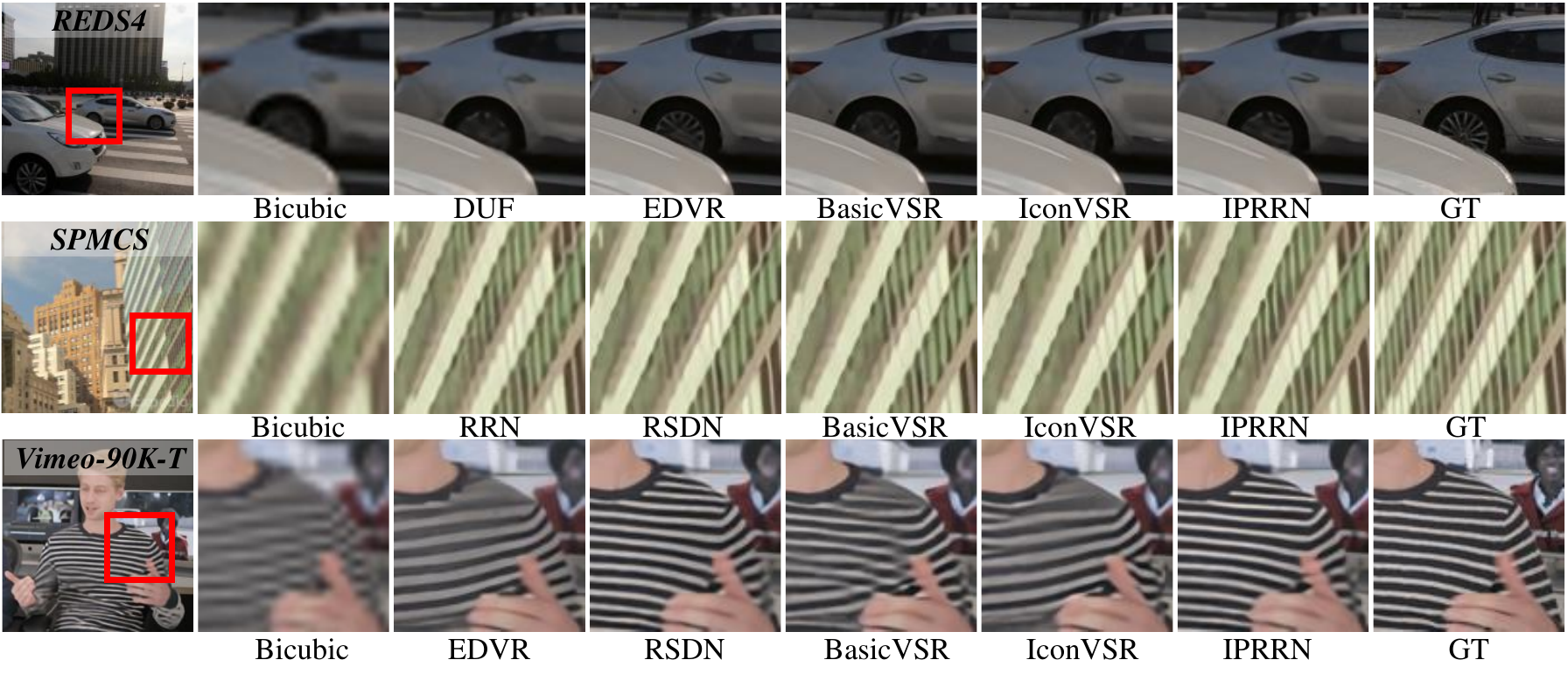}
\caption{Super-resolution qualitative comparison with sampling factor $\times$4 on REDS4 \cite{nah2019ntire}, SPMCS \cite{tao2017detail}and Vimeo-90K-T \cite{xue2019video}. Zoom in for better visualization.}
\end{figure*}

\begin{figure*}[t]  
\centering
\includegraphics[width=7.1in]{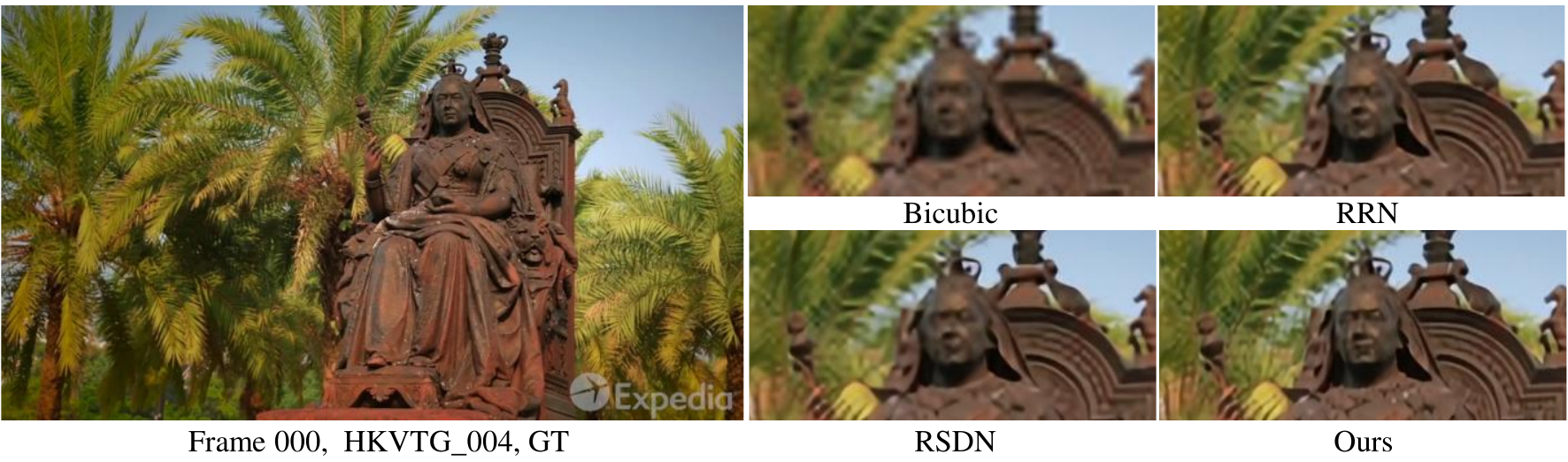}
\caption{Super-resolution qualitative comparison with other unidirectional recurrent methods on the frame of HKVTG\_004 on SPMCS \cite{tao2017detail}. Zoom in for better visualization.}
\end{figure*}

\subsubsection{The number of residual dense blocks in RRNet}
This section further probes into impacts of different numbers of RDB in RRNet on reconstruction performance. As we can see from Table \uppercase\expandafter{\romannumeral6}, recovery performance improves as the number of RDBs increases. The results show that the increase of the number of RDB can make the extracted low frequency and high frequency information more accurate, which improves the performance of reconstruction. It is worth noting  that while performance can be further improved by using more blocks, the time it takes to recover will also increase accordingly.

\subsection{Comparison with State-of-the-Arts}
We compared our approach to the latest VSR methods, inclincluding Bicubic, FRVSR \cite{sajjadi2018frame}, DUF \cite{jo2018deep}, RBPN \cite{haris2019recurrent}, PFNL \cite{yi2019progressive}, EDVR \cite{wang2019edvr}, TGA \cite{isobe2020temporal}, RLSP \cite{song2021multi}, RSDN \cite{isobe2020video}, RRN \cite{isobe2020revisiting}, and BasicVSR\&IconVSR \cite{chan2021basicvsr}. Due to serious boundary effect, the DUF method needs to crop 8 pixels near the image boundary, which will lead to an unbalanced evaluation of the model. There is no added restriction in our method, that is, all pixels from the first frame to the last frame are both involved in the evaluation. Since our approach is a combination of sliding window and unidirectional recurrent, it is unfair and meaningless to compare it with bidirectional methods. Such methods require modeling on the entire sequence, they are not capable for many tasks with low latency requirements, and they need to cache a large number of intermediate features, making it difficult to deploy on resource-constrained devices.

\subsubsection{Quantitative evaluation}
Table \uppercase\expandafter{\romannumeral2} lists the quantitative results obtained on Vid4 \cite{liu2011bayesian}, and Table \uppercase\expandafter{\romannumeral7} lists the quantitative results obtained on REDS4 \cite{nah2019ntire}, SPMCS \cite{tao2017detail}, Vimeo-90K-T \cite{xue2019video} datasets.

\paragraph{Vid4 test set evaluation}
Vid4 \cite{liu2011bayesian} is a widely used baseline test set consisting of four video sequences: calendar, city, foliage, and walk. Table \uppercase\expandafter{\romannumeral2} lists the PSNR and SSIM results on Vid4 \cite{liu2011bayesian} compared with other methods. The results show that our method is superior to all other methods in PSNR and SSIM index, including sliding window method, bidirectional recurrent method and unidirectional recurrent method. Specifically, IPRRN is 0.40dB and 0.35dB higher than the state-of-the-art method BasicVSR and IconVSR \cite{chan2021basicvsr} in the PSNR evaluation index. BasicVSR and IconVSR \cite{chan2021basicvsr} are the offline super-resolution models, it is unfair to make comparison between  our method and them. However, it can be seen from the quantitative evaluation results that our method is better than them in all indicators of application scenarios, parameters, floating point operations and PSNR. The remarkable thing is that our method performs very well on the city scene, with 1.32dB higher than the second place, which helps us stand out from these methods.

\paragraph{SPMCS and REDS4 test set evaluation}
We then tested our method on the SPMCS \cite{tao2017detail} test set. Compared with Vid4 \cite{liu2011bayesian}, SPMCS \cite{tao2017detail} contains more high-frequency information with higher resolution, which requires algorithms with strong recovery ability. As is shown in Table \uppercase\expandafter{\romannumeral7},  our model achieves the best results in both PSNR and SSIM. In particular, our model has better performance with 0.24 dB and 0.28 dB of PSNR higher than those of IconVSR and BasicVSR \cite{chan2021basicvsr} respectively. By comparison with other methods on REDS4 \cite{nah2019ntire}, we extend our training sequences to 15 to facilitate longer time-dependent learning. As is shown in Table \uppercase\expandafter{\romannumeral7}, we surpass all of the sliding window models in performance and runtime,  with even 6 times faster than EDVR \cite{wang2019edvr}.

\paragraph{Vimeo-90K-T test set evaluation}
Vimeo-90k-T \cite{xue2019video} is a special test set because it has only 7 frames in each video. Only the center frame is expected to be restored, which impedes a fair comparison with our recurrent method. One of the advantages of unidirectional recurrent method is the ability of long-term modeling. However, the length of video does not cover the length of the temporal receptive field, which gives limits to the capabilities of our method. As is shown in Table \uppercase\expandafter{\romannumeral7}, our performance outperformed all unidirectional recurrent methods and comparable to BasicVSR \cite{chan2021basicvsr}, despite BasicVSR \cite{chan2021basicvsr} is a bidirectional recurrent method. Since the sliding window method is more capable of reconstruction of short-term sequences, EDVR \cite{wang2019edvr} and TGA \cite{isobe2020temporal}, based on the sliding window method, enjoys better performance with higher PSNR in the evaluation of Vimeo-90K-T \cite{xue2019video}. 

\subsubsection{Qualitative evaluation}
Figure 7 illustrates the qualitative results of two scenarios from the Vid4 \cite{liu2011bayesian} test set. As we can see from the magnified area, our method can restore more reliable and finer details. In the example of the calendar video, our method can recover English words closer to the ground truth than other methods. The city video shows that IPRRN is far superior to current state-of-the-art methods in restoring details of buildings, and only our method can restore the correct texture. The qualitative comparison with other state-of-the-art methods is shown in Figure 8. It can be observed that previous methods are either prone to generate some artifacts (e.g., wrong stripes on clothes in Vimeo-90K-T \cite{xue2019video}) or unable to recover missing details (e.g., the details of automobile window edge in SPMCS \cite{tao2017detail}). By contrast, the outputs of IPRRN have edge texture with higher quality and present finer internal details. Figure 9 shows the super-resolution qualitative comparison with other unidirectional recurrent methods on the frame of HKVTG\_004 on SPMCS \cite{tao2017detail}. Our method restores better facial detail texture in the first frame, which is due to the compensation of temporal receptive field in IPNet and the reuse of hierarchical features in RRNet.

\section{Conclusion}
In this paper, we propose a new information prebuilt recurrent reconstruction network for VSR. To balance the temporal receptive field, we input the useful information in earlier video into the information prebuilt network to build the initial hidden state. For more efficient reconstruction, we add residual dense blocks to the recurrent architecture to refine features. The combination of initial hidden state and recurrent reconstruction network allows us to succeed in video recovery with high efficiency and time consistency. Extensive experiments have shown that our network achieves state-of-the-art performance across multiple VSR benchmark test sets. In the future, we plan to enhance the overall video recovery effect by enhancing the propagation relevance and optimizing the structural components to devise a more accurate and efficient VSR method.

\ifCLASSOPTIONcaptionsoff
  \newpage
\fi



% http://www.michaelshell.org/tex/ieeetran/bib tex/
%\bibliographystyle{IEEEtran}
%\begin{thebibliography}{1} 
  % IEEEtran
\bibliographystyle{IEEEtranN}
\bibliography{IEEEabrv, mylib}

% Generated by IEEEtranN.bst, version: 1.14 (2015/08/26)
\begin{thebibliography}{41}
\providecommand{\natexlab}[1]{#1}
\providecommand{\url}[1]{#1}
\csname url@samestyle\endcsname
\providecommand{\newblock}{\relax}
\providecommand{\bibinfo}[2]{#2}
\providecommand{\BIBentrySTDinterwordspacing}{\spaceskip=0pt\relax}
\providecommand{\BIBentryALTinterwordstretchfactor}{4}
\providecommand{\BIBentryALTinterwordspacing}{\spaceskip=\fontdimen2\font plus
\BIBentryALTinterwordstretchfactor\fontdimen3\font minus
  \fontdimen4\font\relax}
\providecommand{\BIBforeignlanguage}[2]{{%
\expandafter\ifx\csname l@#1\endcsname\relax
\typeout{** WARNING: IEEEtranN.bst: No hyphenation pattern has been}%
\typeout{** loaded for the language `#1'. Using the pattern for}%
\typeout{** the default language instead.}%
\else
\language=\csname l@#1\endcsname
\fi
#2}}
\providecommand{\BIBdecl}{\relax}
\BIBdecl

\bibitem[Dong et~al.(2015)Dong, Loy, He, and Tang]{dong2015image}
C.~Dong, C.~C. Loy, K.~He, and X.~Tang, ``Image super-resolution using deep
  convolutional networks,'' \emph{IEEE Trans. Pattern Anal. Mach. Intell.},
  vol.~38, no.~2, pp. 295--307, 2015.

\bibitem[Shi et~al.(2016)]{shi2016real}
W.~Shi \emph{et~al.}, ``Real-time single image and video super-resolution using
  an efficient sub-pixel convolutional neural network,'' in \emph{Proc. IEEE
  Conf. Comput. Vis. Pattern Recognit.}, 2016, pp. 1874--1883.

\bibitem[Ledig et~al.(2017)]{ledig2017photo}
C.~Ledig \emph{et~al.}, ``Photo-realistic single image super-resolution using a
  generative adversarial network,'' in \emph{Proc. IEEE Conf. Comput. Vis.
  Pattern Recognit.}, 2017, pp. 4681--4690.

\bibitem[Niu et~al.(2020)]{niu2020single}
B.~Niu \emph{et~al.}, ``Single image super-resolution via a holistic attention
  network,'' in \emph{Proc. Eur. Conf. Comput. Vis.}\hskip 1em plus 0.5em minus
  0.4em\relax Springer, 2020, pp. 191--207.

\bibitem[Wang et~al.(2021)]{wang2021unsupervised}
L.~Wang \emph{et~al.}, ``Unsupervised degradation representation learning for
  blind super-resolution,'' in \emph{Proc. IEEE/CVF Conf. Comput. Vis. Pattern
  Recognit.}, 2021, pp. 10\,581--10\,590.

\bibitem[Menon et~al.(2020)]{menon2020pulse}
S.~Menon \emph{et~al.}, ``Pulse: Self-supervised photo upsampling via latent
  space exploration of generative models,'' in \emph{Proc. IEEE/CVF Conf.
  Comput. Vis. Pattern Recognit.}, 2020, pp. 2437--2445.

\bibitem[Yi et~al.(2019)Yi, Wang, Jiang, Jiang, and Ma]{yi2019progressive}
P.~Yi, Z.~Wang, K.~Jiang, J.~Jiang, and J.~Ma, ``Progressive fusion video
  super-resolution network via exploiting non-local spatio-temporal
  correlations,'' in \emph{Proc. IEEE/CVF IEEE Int. Conf. Comput. Vis.}, 2019,
  pp. 3106--3115.

\bibitem[Tian et~al.(2020)Tian, Zhang, Fu, and Xu]{tian2020tdan}
Y.~Tian, Y.~Zhang, Y.~Fu, and C.~Xu, ``Tdan: Temporally-deformable alignment
  network for video super-resolution,'' in \emph{Proc. IEEE/CVF Conf. Comput.
  Vis. Pattern Recognit.}, 2020, pp. 3360--3369.

\bibitem[Kappeler et~al.(2016)Kappeler, Yoo, Dai, and
  Katsaggelos]{kappeler2016video}
A.~Kappeler, S.~Yoo, Q.~Dai, and A.~K. Katsaggelos, ``Video super-resolution
  with convolutional neural networks,'' \emph{IEEE Trans. Comput. Imag.},
  vol.~2, no.~2, pp. 109--122, 2016.

\bibitem[Li et~al.(2020{\natexlab{a}})Li, Bai, and Zhao]{li2020learning}
F.~Li, H.~Bai, and Y.~Zhao, ``Learning a deep dual attention network for video
  super-resolution,'' \emph{IEEE Trans. Image Process.}, vol.~29, pp.
  4474--4488, 2020.

\bibitem[Yan et~al.(2019)Yan, Lin, and Tan]{yan2019frame}
B.~Yan, C.~Lin, and W.~Tan, ``Frame and feature-context video
  super-resolution,'' in \emph{Proc. AAAI Conf. Artif. Intell.}, vol.~33,
  no.~01, 2019, pp. 5597--5604.

\bibitem[Chu et~al.(2020)Chu, Xie, Mayer, Leal-Taix{\'e}, and
  Thuerey]{chu2020learning}
M.~Chu, Y.~Xie, J.~Mayer, L.~Leal-Taix{\'e}, and N.~Thuerey, ``Learning
  temporal coherence via self-supervision for gan-based video generation,''
  \emph{ACM Trans. Graph.}, vol.~39, no.~4, pp. 75--1, 2020.

\bibitem[Isobe et~al.(2020{\natexlab{a}})]{isobe2020temporal}
T.~Isobe \emph{et~al.}, ``Video super-resolution with temporal group
  attention,'' in \emph{Proc. IEEE/CVF Conf. Comput. Vis. Pattern Recognit.},
  2020, pp. 8008--8017.

\bibitem[Li et~al.(2020{\natexlab{b}})Li, Tao, Guo, Qi, Lu, and
  Jia]{li2020mucan}
W.~Li, X.~Tao, T.~Guo, L.~Qi, J.~Lu, and J.~Jia, ``Mucan: Multi-correspondence
  aggregation network for video super-resolution,'' in \emph{Proc. Eur. Conf.
  Comput. Vis.}\hskip 1em plus 0.5em minus 0.4em\relax Springer, 2020, pp.
  335--351.

\bibitem[Xiang et~al.(2020)Xiang, Tian, Zhang, Fu, Allebach, and
  Xu]{xiang2020zooming}
X.~Xiang, Y.~Tian, Y.~Zhang, Y.~Fu, J.~P. Allebach, and C.~Xu, ``Zooming
  slow-mo: Fast and accurate one-stage space-time video super-resolution,'' in
  \emph{Proc. IEEE/CVF Conf. Comput. Vis. Pattern Recognit.}, 2020, pp.
  3370--3379.

\bibitem[Kang et~al.(2020)Kang, Jo, Oh, Vajda, and Kim]{kang2020deep}
J.~Kang, Y.~Jo, S.~W. Oh, P.~Vajda, and S.~J. Kim, ``Deep space-time video
  upsampling networks,'' in \emph{Proc. Eur. Conf. Comput. Vis.}\hskip 1em plus
  0.5em minus 0.4em\relax Springer, 2020, pp. 701--717.

\bibitem[Xu et~al.(2021)Xu, Xu, Li, Wang, Sun, and Cheng]{xu2021temporal}
G.~Xu, J.~Xu, Z.~Li, L.~Wang, X.~Sun, and M.-M. Cheng, ``Temporal modulation
  network for controllable space-time video super-resolution,'' in \emph{Proc.
  IEEE/CVF Conf. Comput. Vis. Pattern Recognit.}, 2021, pp. 6388--6397.

\bibitem[Sajjadi et~al.(2018)Sajjadi, Vemulapalli, and Brown]{sajjadi2018frame}
M.~S. Sajjadi, R.~Vemulapalli, and M.~Brown, ``Frame-recurrent video
  super-resolution,'' in \emph{Proc. IEEE Conf. Comput. Vis. Pattern
  Recognit.}, 2018, pp. 6626--6634.

\bibitem[Fuoli et~al.(2019)Fuoli, Gu, and Timofte]{fuoli2019efficient}
D.~Fuoli, S.~Gu, and R.~Timofte, ``Efficient video super-resolution through
  recurrent latent space propagation,'' in \emph{Proc. IEEE Int. Conf. Comput.
  Vis. Workshop}.\hskip 1em plus 0.5em minus 0.4em\relax IEEE, 2019, pp.
  3476--3485.

\bibitem[Isobe et~al.(2020{\natexlab{b}})Isobe, Zhu, Jia, and
  Wang]{isobe2020revisiting}
T.~Isobe, F.~Zhu, X.~Jia, and S.~Wang, ``Revisiting temporal modeling for video
  super-resolution,'' \emph{Proc. The British Mach. Vis. Conf.}, 2020.

\bibitem[Isobe et~al.(2020{\natexlab{c}})Isobe, Jia, Gu, Li, Wang, and
  Tian]{isobe2020video}
T.~Isobe, X.~Jia, S.~Gu, S.~Li, S.~Wang, and Q.~Tian, ``Video super-resolution
  with recurrent structure-detail network,'' in \emph{Proc. Eur. Conf. Comput.
  Vis.}\hskip 1em plus 0.5em minus 0.4em\relax Springer, 2020, pp. 645--660.

\bibitem[Huang et~al.(2017)Huang, Wang, and Wang]{huang2017video}
Y.~Huang, W.~Wang, and L.~Wang, ``Video super-resolution via bidirectional
  recurrent convolutional networks,'' \emph{IEEE Trans. Pattern Anal. Mach.
  Intell.}, vol.~40, no.~4, pp. 1015--1028, 2017.

\bibitem[Li et~al.(2019)Li, He, Du, Zhang, Xu, and Tao]{li2019fast}
S.~Li, F.~He, B.~Du, L.~Zhang, Y.~Xu, and D.~Tao, ``Fast spatio-temporal
  residual network for video super-resolution,'' in \emph{Proc. IEEE/CVF Conf.
  Comput. Vis. Pattern Recognit.}, 2019, pp. 10\,522--10\,531.

\bibitem[Chan et~al.(2021)Chan, Wang, Yu, Dong, and Loy]{chan2021basicvsr}
K.~C. Chan, X.~Wang, K.~Yu, C.~Dong, and C.~C. Loy, ``Basicvsr: The search for
  essential components in video super-resolution and beyond,'' in \emph{Proc.
  IEEE/CVF Conf. Comput. Vis. Pattern Recognit.}, 2021, pp. 4947--4956.

\bibitem[Zhu et~al.(2019)]{zhu2019residual}
X.~Zhu \emph{et~al.}, ``Residual invertible spatio-temporal network for video
  super-resolution,'' in \emph{Proc. AAAI Conf. Artif. Intell.}, vol.~33,
  no.~01, 2019, pp. 5981--5988.

\bibitem[Wang et~al.(2018)Wang, Guo, Lin, Deng, and An]{wang2018learning}
L.~Wang, Y.~Guo, Z.~Lin, X.~Deng, and W.~An, ``Learning for video
  super-resolution through hr optical flow estimation,'' in \emph{Proc. Asia
  Conf. Comput. Vis.}\hskip 1em plus 0.5em minus 0.4em\relax Springer, 2018,
  pp. 514--529.

\bibitem[Caballero et~al.(2017)]{caballero2017real}
J.~Caballero \emph{et~al.}, ``Real-time video super-resolution with
  spatio-temporal networks and motion compensation,'' in \emph{Proc. IEEE Conf.
  Comput. Vis. Pattern Recognit.}, 2017, pp. 4778--4787.

\bibitem[He et~al.(2016)He, Zhang, Ren, and Sun]{he2016deep}
K.~He, X.~Zhang, S.~Ren, and J.~Sun, ``Deep residual learning for image
  recognition,'' in \emph{Proc. IEEE Conf. Comput. Vis. Pattern Recognit.},
  2016, pp. 770--778.

\bibitem[Zhang et~al.(2018{\natexlab{a}})Zhang, Tian, Kong, Zhong, and
  Fu]{zhang2018residual}
Y.~Zhang, Y.~Tian, Y.~Kong, B.~Zhong, and Y.~Fu, ``Residual dense network for
  image super-resolution,'' in \emph{Proc. IEEE Conf. Comput. Vis. Pattern
  Recognit.}, 2018, pp. 2472--2481.

\bibitem[Jo et~al.(2018)Jo, Oh, Kang, and Kim]{jo2018deep}
Y.~Jo, S.~W. Oh, J.~Kang, and S.~J. Kim, ``Deep video super-resolution network
  using dynamic upsampling filters without explicit motion compensation,'' in
  \emph{Proc. IEEE Conf. Comput. Vis. Pattern Recognit.}, 2018, pp. 3224--3232.

\bibitem[Wang et~al.(2019)Wang, Chan, Yu, Dong, and Change~Loy]{wang2019edvr}
X.~Wang, K.~C. Chan, K.~Yu, C.~Dong, and C.~Change~Loy, ``Edvr: Video
  restoration with enhanced deformable convolutional networks,'' in \emph{Proc.
  IEEE Conf. Comput. Vis. Pattern Recognit. Workshops}, 2019, pp. 0--0.

\bibitem[Zhang et~al.(2018{\natexlab{b}})Zhang, Li, Li, Wang, Zhong, and
  Fu]{zhang2018image}
Y.~Zhang, K.~Li, K.~Li, L.~Wang, B.~Zhong, and Y.~Fu, ``Image super-resolution
  using very deep residual channel attention networks,'' in \emph{Proc. Eur.
  Conf. Comput. Vis.}, 2018, pp. 286--301.

\bibitem[Hu et~al.(2018)Hu, Shen, and Sun]{hu2018squeeze}
J.~Hu, L.~Shen, and G.~Sun, ``Squeeze-and-excitation networks,'' in \emph{Proc.
  IEEE Conf. Comput. Vis. Pattern Recognit.}, 2018, pp. 7132--7141.

\bibitem[Xue et~al.(2019)Xue, Chen, Wu, Wei, and Freeman]{xue2019video}
T.~Xue, B.~Chen, J.~Wu, D.~Wei, and W.~T. Freeman, ``Video enhancement with
  task-oriented flow,'' \emph{International Journal of Computer Vision}, vol.
  127, no.~8, pp. 1106--1125, 2019.

\bibitem[Nah et~al.(2019)Nah, Baik, Hong, Moon, Son, Timofte, and
  Mu~Lee]{nah2019ntire}
S.~Nah, S.~Baik, S.~Hong, G.~Moon, S.~Son, R.~Timofte, and K.~Mu~Lee, ``Ntire
  2019 challenge on video deblurring and super-resolution: Dataset and study,''
  in \emph{Proc. IEEE Conf. Comput. Vis. Pattern Recognit. Workshops}, 2019,
  pp. 0--0.

\bibitem[Liu and Sun(2011)]{liu2011bayesian}
C.~Liu and D.~Sun, ``A bayesian approach to adaptive video super resolution,''
  in \emph{Proc. IEEE Conf. Comput. Vis. Pattern Recognit.}\hskip 1em plus
  0.5em minus 0.4em\relax IEEE, 2011, pp. 209--216.

\bibitem[Tao et~al.(2017)Tao, Gao, Liao, Wang, and Jia]{tao2017detail}
X.~Tao, H.~Gao, R.~Liao, J.~Wang, and J.~Jia, ``Detail-revealing deep video
  super-resolution,'' in \emph{Proc. IEEE Int. Conf. Comput. Vis.}, 2017, pp.
  4472--4480.

\bibitem[Wang et~al.(2004)Wang, Bovik, Sheikh, and Simoncelli]{wang2004image}
Z.~Wang, A.~C. Bovik, H.~R. Sheikh, and E.~P. Simoncelli, ``Image quality
  assessment: from error visibility to structural similarity,'' \emph{IEEE
  Trans. Image Process.}, vol.~13, no.~4, pp. 600--612, 2004.

\bibitem[Kingma and Ba(2014)]{kingma2014adam}
D.~P. Kingma and J.~Ba, ``Adam: A method for stochastic optimization,''
  \emph{arXiv preprint arXiv:1412.6980}, 2014.

\bibitem[Haris et~al.(2019)Haris, Shakhnarovich, and Ukita]{haris2019recurrent}
M.~Haris, G.~Shakhnarovich, and N.~Ukita, ``Recurrent back-projection network
  for video super-resolution,'' in \emph{Proc. IEEE/CVF Conf. Comput. Vis.
  Pattern Recognit.}, 2019, pp. 3897--3906.

\bibitem[Song et~al.(2021)]{song2021multi}
H.~Song \emph{et~al.}, ``Multi-stage feature fusion network for video
  super-resolution,'' \emph{IEEE Trans. Image Process.}, vol.~30, pp.
  2923--2934, 2021.

\end{thebibliography}
%\end{thebibliography}


\begin{IEEEbiography}[{\includegraphics[width=1in,height=1.25in,clip,keepaspectratio]{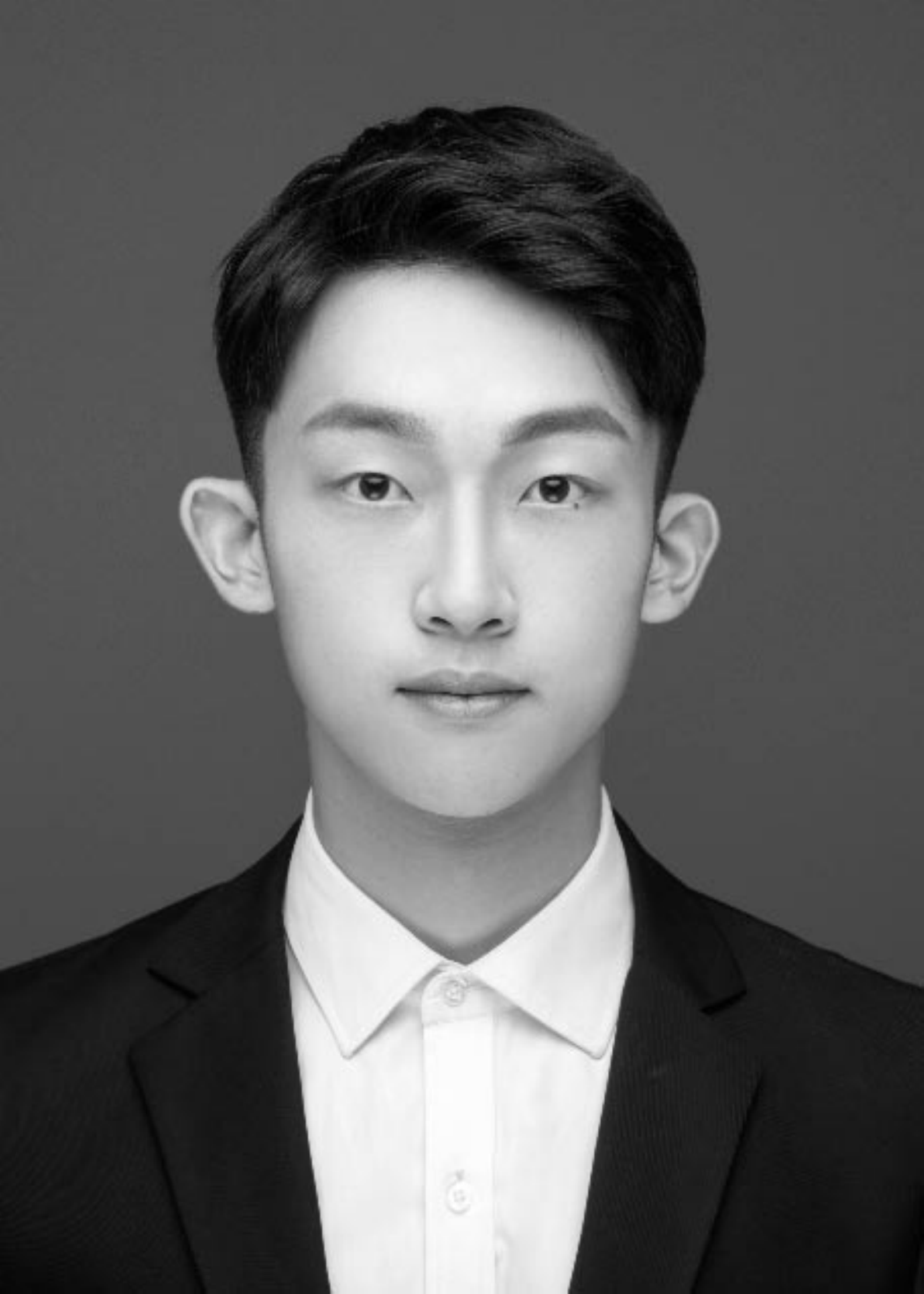}}]{Shuyun Wang}
  received the B.S. degree from Hebei Normal University, China, in 2019. He is currently pursuing the M.S. degree with the School of Artificial Intelligence, Hebei University of Technology, China. His research interests are in image/video super-resolution, video Frame interpolation and other low-level computer vision tasks.
  \end{IEEEbiography}
  % \vspace{-10 mm}

\begin{IEEEbiography}[{\includegraphics[width=1in,height=1.25in,clip,keepaspectratio]{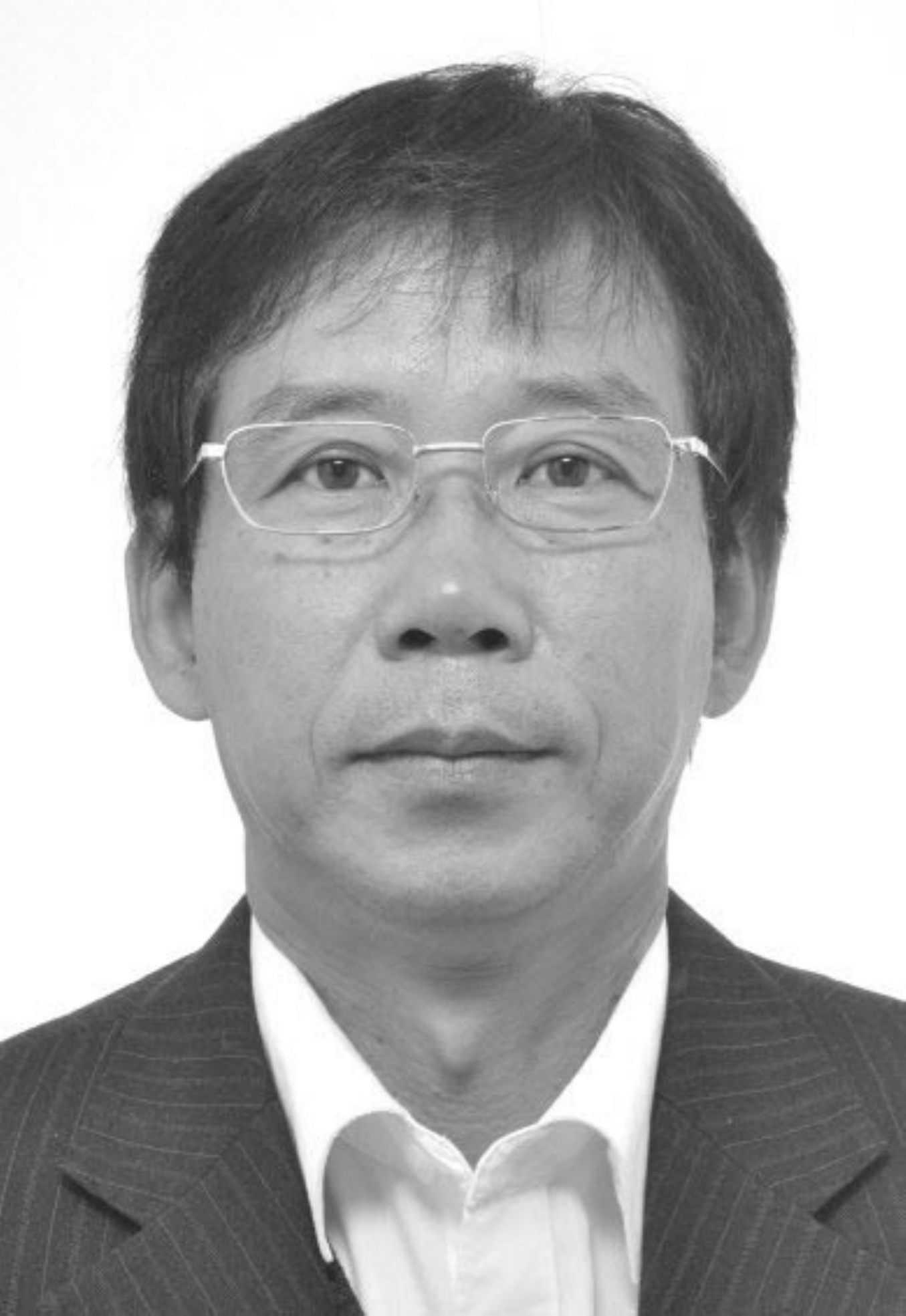}}]{Ming Yu}
  received the M.Sc. degree in communication Electronic System from Hebei University of Technology in 1989 and the Ph.D. degree in communication and information system from Beijing Institute of Technology, China, in 1999. He is currently a professor and he is a director of the China National Information and Electronics Postgraduate Education Committee and the deputy director of the Hebei Machine Learning Society. His research interests include Biometric recognition based on the fusion of speech and image visual information, Video data mining, face recognition, handwriting identification system and image processing.
  \end{IEEEbiography}
% \vspace{-10 mm}

\begin{IEEEbiography}[{\includegraphics[width=1in,height=1.25in,clip,keepaspectratio]{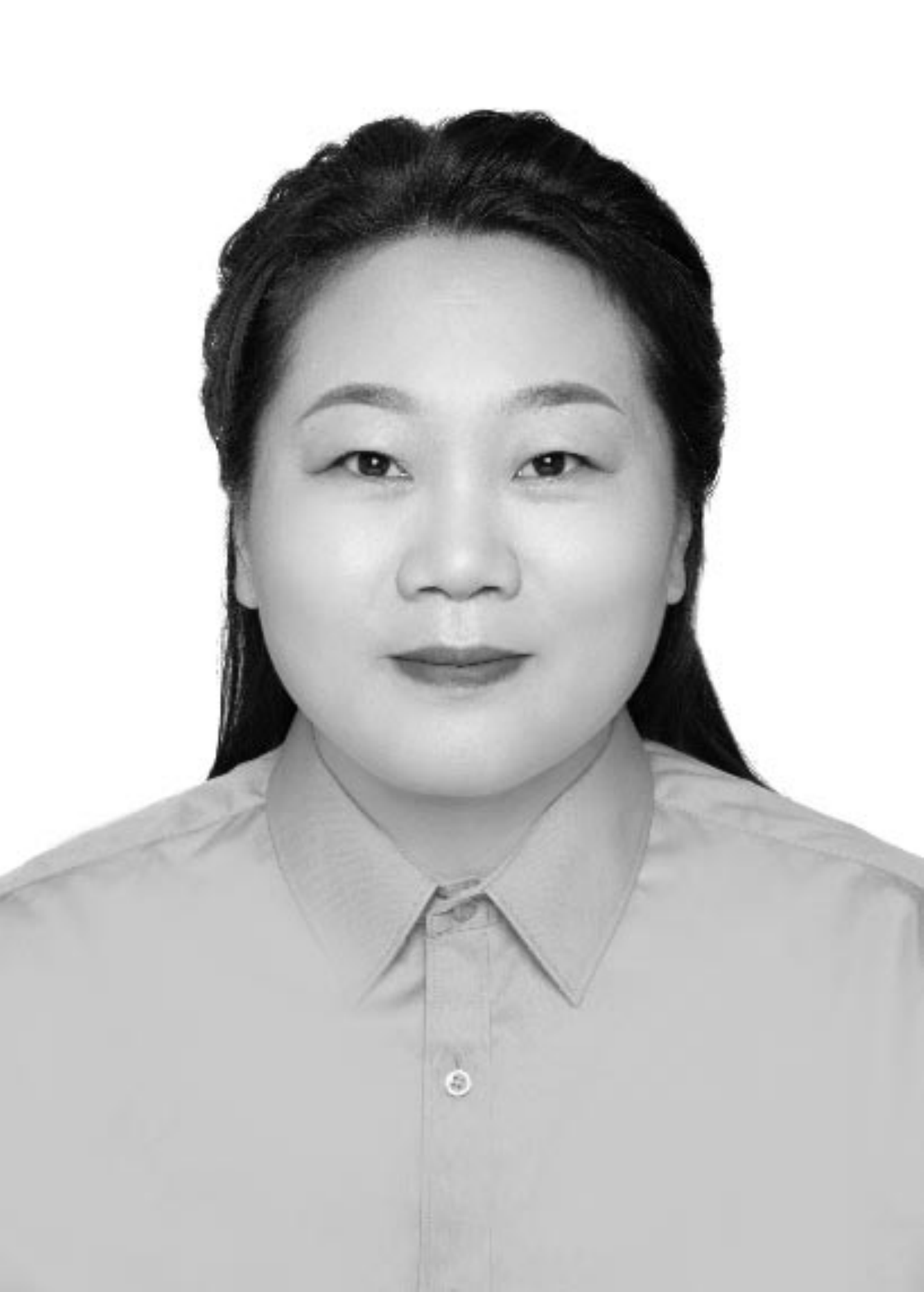}}]{Cuihong Xue}
  received the M.Sc. degree in received the B.E. degree in computer science and technology from Hebei University of Technology, China, in 2007, and the Ph.D. degree in microelectronics and solid state electronics from Hebei University of Technology, China, in 2012. He is currently a teacher at Tianjin University of technology. Her research interests include sign language recognition, face recognition, object recognition, supper resolution reconstruction and image processing.
  \end{IEEEbiography}
% \vspace{-10 mm}

\begin{IEEEbiography}[{\includegraphics[width=1in,height=1.25in,clip,keepaspectratio]{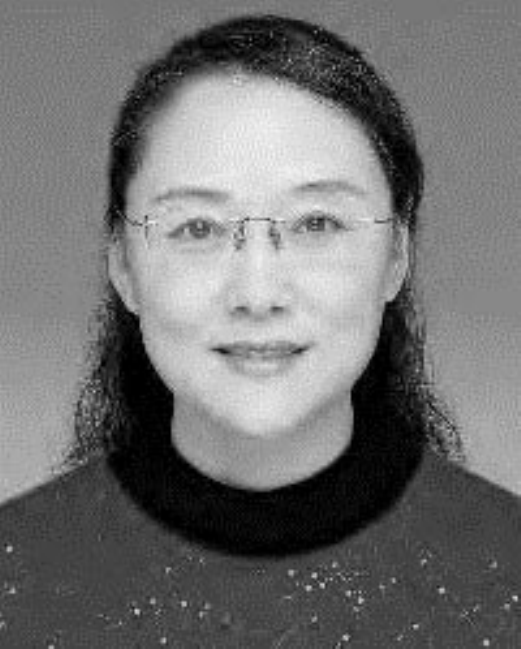}}]{Yingchun Guo}
received the Ph.D. degree from the School of Information, Tianjin University, Tianjin, China, in 2006. She is currently an Associate Professor with the School of Artificial Intelligence, Hebei University of Science and Technology, Tianjin. Her research interests include image saliency and its application, image processing, artificial intelligence, and image compression.
\end{IEEEbiography}
% \vspace{-10 mm}

\begin{IEEEbiography}[{\includegraphics[width=1in,height=1.25in,clip,keepaspectratio]{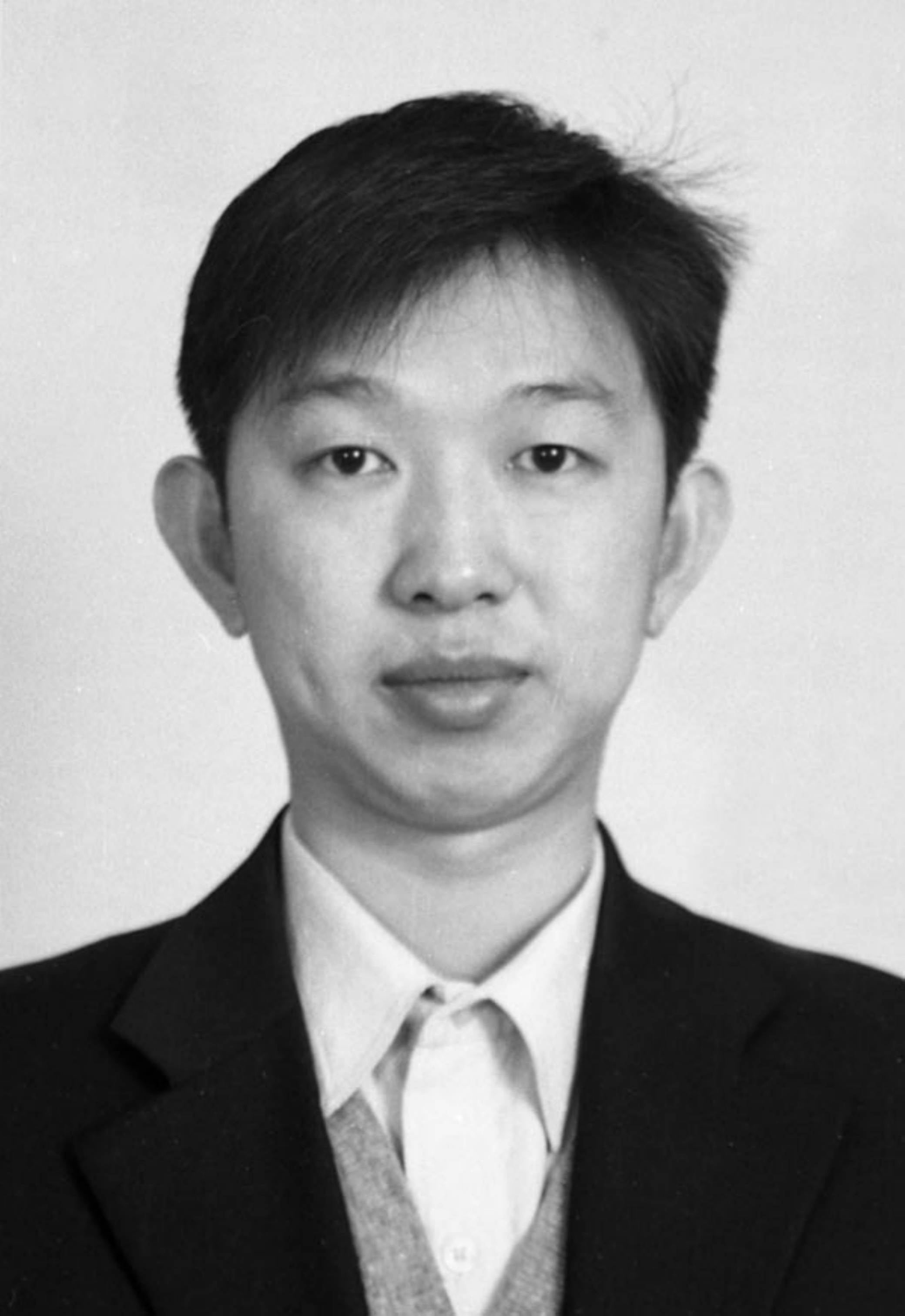}}]{Gang Yan}
received the B.S. and M.S. degrees in computer science, and the Ph.D. degree in microelectronics and solid electronics from the Hebei University of Technology, Tianjin, China, in 2000, 2003, and 2019, respectively. He is currently an Associate Professor with the School of Artificial Intelligence, Hebei University of Technology. His current research interests include intelligent transport systems and mobile Internet application development.
\end{IEEEbiography}

%\newpage


%\vfill


\end{document}